\documentclass [11pt,twoside]{article}

\usepackage{shrthnds}
\usepackage{cite}
\usepackage{graphicx}
\usepackage[usenames]{color}
\usepackage{subfigure}
\usepackage{setspace}
\usepackage{multirow}
\usepackage[hang,small]{caption}
\usepackage{amsmath,amssymb}
\usepackage{slashed}                  
\usepackage{geometry}
\usepackage{fancyhdr}
\usepackage{titlesec,titletoc}
  \titleformat{\section}{\Large\sf\bfseries}{\thesection}{1em}{}
  \titleformat{\subsection}{\large\sf\bfseries}{\thesubsection}{1em}{}
\newcommand{\auth}[2]{{\large #1}\footnote{email:~#2}}
\newcommand{\affS}[1]{{\it #1}}
\newcommand{\verttle}[2][~]{\vspace{-1cm}\begin{flushright}{\small #1}\end{flushright}\vspace{0.5cm} {\sf\bfseries #2}}
\newcommand{\nabstract}[2][]{\bc\begin{minipage}{0.9\textwidth}\begin{spacing}{1}{\small {\sf\bfseries Abstract:} #2 }\end{spacing}
#1 \end{minipage}\ec}
\newcommand{\nkeywords}[1][]{~\\\small{ {\sf\bfseries Keywords:} #1 }}

\title{\verttle{Associated Production of a KK-Graviton with a Higgs Boson via Gluon Fusion at the LHC\\}}
\author{
  \auth{Ambresh Shivaji}{ambresh@iopb.res.in},
  \auth{Subhadip Mitra}{smitra@iopb.res.in}~and
  \auth{Pankaj Agrawal}{agrawal@iopb.res.in}\\
  \affS{ Institute of Physics, Saink School Post, Bhubaneswar 751 005, India}
}
\newcommand{\pghdr}{\footnotesize A. Shivaji {\it et al.} -- Associated Production of a KK-Graviton and a Higgs \ldots}
\date{March 11, 2012}

\begin{document}

\maketitle
\nabstract[\nkeywords{
ADD model, RS model, Graviton production at LHC, Higgs, Top quark loop, heavy quark decoupling
}]{
In order to solve the hierarchy problem, several extra-dimensional models have received considerable 
attention. We have considered a process where a Higgs boson is produced in association 
with a KK-graviton ($G_{\rm KK}$) at the LHC. At the leading order, this process occurs through gluon fusion mechanism 
$gg \to h G_{\rm KK}$ via a quark loop. We compute the cross section and examine some features of this
process in the ADD model. We find that the quark in the loop does not decouple in the large quark-mass limit 
just as in the case of $gg\to h$ process. We compute the cross section of this process 
for the case of the RS model also. We examine the feasibility of this process being observed at the LHC.
}
\bigskip

\section{Introduction}\label{sec:introduction}
The standard model (SM) has been very successful in explaining a wide array of experimental data\cite{Nakamura:2010zzi}.
However, it remains incomplete and a number of different approaches have been taken to address
associated issues\cite{jellis,Lykken:2010mc}. One such approach is the extra-dimensional extensions of the SM\cite{
ADD,RS1,RS2,Antoniadis,Appelquist}. These models were 
motivated due to the possibility of experimental access to extra dimensions and solving 
the hierarchy problem\cite{Davoudiasl:1999jd,Han:1998sg,Giudice:1998ck,Mirabelli:1998rt,Ghosh}. 
In these models, the number of spacial dimensions 
is assumed to be $3+d$ where $d$ is the number
of extra dimensions. These extra dimensions are generally assumed to be compactified. 
Gravity can freely propagate through the extra-dimensions
but, depending on the model, the SM fields can either propagate in the bulk or live 
on a boundary of the bulk ($3$-brane). Moreover, these models differ over the number and 
the size of the extra-dimensions. We consider two such models. In one case $d\geq 2$ and 
the scale of the extra-dimensions can be as large as a $\m m$, while for the 
other $d=1$ and the scale of the extra-dimension is slightly larger than the Planck length.

In the model proposed by Arkani-Hamed, Dimopoulos and Dvali (ADD)\cite{ADD}, all the SM fields, unlike gravity, 
are confined to a $3$-brane. Due to the compactification of the extra-dimensions, one can relate the full 
$4+d$ dimensional theory of gravity to a 4 dimensional theory of gravity by a Kaluza-Klein (KK) reduction 
\cite{Kaluza,Klein,Han:1998sg,Giudice:1998ck}. In the 4 dimensional theory, gravity appears as an infinite tower of
gravity fields with increasing mass (KK-modes) starting from zero rather than a single massless graviton 
field. Following Ref. \cite{Han:1998sg}, we assume a $4+d$ dimensional bulk with the extra-dimensions 
compactified on a $d$-dimensional torus of radius $R/2\pi$ for simplicity.The masses of the different 
KK-modes of the graviton are given by,
\ba
M_{\vec n}^2 \equiv M_{\{n_1, n_2,\, \ldots,\, n_d\}}^2 = \frac{4\pi^2 }{R^2}\sum_{i=1}^d \left(n_i ^2\right), \label{eq:massKK}
\ea
where each $n_i = \{0,1,2, \ldots\}$. This 4 dimensional theory of gravity is treated as an effective theory valid below some scale, $M_{\rm S}$. 
This scale of the extra-dimensional theory, which acts like the ultraviolet cutoff scale of the effective theory, is
related to the Planck constant ($M_{\rm PL}$) as,
\ba
M^{d+2}_{\rm S} = \frac{(4\pi)^{d/2}\G(d/2)}{2R^d} M^2_{\rm PL}.\label{eq:ms}
\ea

The spin-2 component of all the KK modes of the graviton interacts with the SM fields with the gravitational coupling $\kp~(= \sqrt{16\pi G_{\rm N}})$,
\ba
\mc L \supset -\frac{\kp}{2}\sum_{\vec n}  T^{\m\n} G^{\vec n}_{\m\n},
\ea
where $T^{\m\n}$ is the contribution of the SM fields to the energy momentum tensor\cite{Han:1998sg}. The production cross section of a particular KK mode
of the graviton is very small due to the extremely weak nature of the gravitational coupling. However, for large $R$,
 the splitting between two 
successive KK-modes becomes very small and the collective contribution of these closely spaced states can enhance the cross section significantly 
and thereby increase the chance of their detection in the particle colliders. For large $R$, one can make a continuum approximation to 
get the mass density for these states,
\ba
\rho\left(M_{\vec n}^2\right)dM^2_{\vec n} = \frac{R^d M_{\vec n}^{(d-2)}}{(4\pi)^{d/2}\G(d/2)}dM^2_{\vec n}\,.\label{eq:dnsty_states}
\ea
Therefore, one can obtain the cross section for the on-shell KK-graviton production by convoluting the cross section 
for production of a graviton with a fixed mass with the mass density,
\ba
\s = \int \s\left({M^2_{\vec n}}\right) \rho\left(M^2_{\vec n}\right) \, dM^2_{\vec n}\; .
\ea
Due to the density of the KK-graviton modes, there is an enhancement factor of $M^2_{\rm PL}/M_{\rm S}^{d+2}$. The single KK 
mode production cross section, $\s\left({M^2_{\vec n}}\right)$ is suppressed by $1/M^2_{\rm PL}$. Therefore, 
 the cross section, $\s$ is ultimately suppressed only by a factor of $1/M_{\rm S}^{d+2}$. Eq. \ref{eq:ms} 
indicates that depending on $R$, the scale $M_{\rm S}$ can be much 
smaller than the Planck scale. Recent experimental limits on the mass scale are of TeV order ($R \sim \m m$) for $d \ge 2$\cite{Kapner,Chatrchyan}. 
If the mass scale is not far away from the above limit, then it may be possible to see the signatures of this theory in the Large Hadron 
Collider (LHC) at CERN.

Unlike the ADD model, the RS model, proposed by Randall and Sundrum\cite{RS1}, has only one extra-dimension. In this model the 
five dimensional bulk is warped with the following space-time metric,
\ba
ds^2 = \left(e^{-2 k R_c \ph}\right)\et_{\m\n}dx^\m dx^\n + R_c^2d\ph^2,
\ea
where the extra dimension $\ph$ is assumed to be compactified on a $S^1/{\bf Z_2}$ orbifold. The radius of 
compactification is $R_c$ and $k$ is an extra mass scale related to the Planck scale of the 5 dimensional 
theory and the physical Planck scale. At the fixed points of the orbifold, there are two 3-branes -- 
the infrared (IR) or the TeV brane  where the SM fields are localized and the UV or the Planck brane.
Due to the warped nature of the bulk space-time the graviton mass spectrum in this model is quite 
different from that of the ADD model\cite{Goldberger}.
The mass of $n^{\rm th}$ KK mode of the graviton can be written as,
\ba
M_{n} = x_n k {W} \equiv x_n m_0,
\ea
where $x_n$ denotes the $n^{\rm th}$ zero of the Bessel function $J_1(x)$ and $W$ is the warp factor,
\ba
W = e^{-\pi k R_c}.
\ea
In the IR brane, all the massive KK modes of graviton couple with the SM fields with an effective 
gravitational coupling $\kp_{\rm IR}$, given as,
\ba 
\kp_{\rm IR} = \frac{\kp}{W} \equiv \sqrt{2}\frac{c_0}{m_0}.
\ea
However, the massless zeroth mode couples with SM fields with the gravitational coupling $\kp$ and hence its effects can safely be ignored.
The two quantities $c_0 = k/{\bar M}_{\rm PL} = \sqrt{8\pi} k/M_{\rm PL}$ and $m_0 = kW$ are the free parameters of the model. 
To determine $c_0$, we note that the RS model requires the scale $k$ to be smaller than $M_{[5]{\rm PL}}$, 
the 5 dimensional Planck scale. However taking $k$ too small will 
not solve the hierarchy problem. A common choice is to take $ 0.01 \lesssim c_0 \lesssim 0.1$ \cite{Davoudiasl:1999jd,Ghosh,Mathews}. 
The other parameter $m_0$ sets the mass scales of different
KK modes of the graviton. In our calculation we consider only 
the first excited mode of the graviton with mass $M_1 = x_1 m_0 \approx 3.83\, m_0$. Recent experiments 
set the limit on $M_1$ as 1058 GeV (612 GeV) for $c_0 =$ 0.1 (0.01)\cite{Aaltonen:2011xp,Abazov:2010xh}.

Earlier people have looked into production of a spin-2 KK-graviton (referred to as graviton in 
the rest of the paper) in association with the SM vector bosons at the LHC   
\cite{Kumar:2010ca,Kumar:2010kv,ambresh,Gao:2009pn,Karg:2009xk}.
In this paper, we investigate the production of a graviton
in association with the only other elementary boson present in the SM, namely the Higgs boson. At the LHC, a graviton can be produced 
in association with a Higgs boson either 
via gluon fusion or different $q\bar q$ initiated processes. But due to the high $E_{\rm CM}$ of the LHC, 
except for very high
$x$ values, the gluon distribution function (PDF) dominates over all quark distribution functions. 
Apart from the PDF 
suppression, the tree level diagram for a $q\bar q$ initiated process like $q\bar q \to h G_{\rm KK}$ 
have very small cross section. This is because coupling of the light quark to the Higgs boson is
very small and there is a 
zero contribution from the diagrams that involve a Higgs-quark-quark-graviton ($hqqG_{\rm KK}$) vertex; 
this vertex is proportional to the metric, $\et_{\m\n}$, which when contracted with the graviton polarization 
tensor gives zero. The rest of the $qq$ initiated processes mediate via fusions of heavy electroweak vector
bosons whose heavy masses and the small electroweak couplings make the contribution
negligible. Therefore, at the LHC, the gluon fusion mechanism, \emph{i.e.} $gg\to h G_{\rm KK}$ 
is the dominant channel for the production of a Higgs boson in association with a graviton.
This is the mechanism that we consider in this paper.
The paper is organized as follows -- we discuss the associated Higgs and graviton production process
and  present the calculational framework in Section \ref{sec:process}; in Section \ref{sec:results}
we present our results and then finally in Section \ref{sec:conclusions} we give our conclusions.

\section{The Process}\label{sec:process}

\begin{figure}[!ht]
\bc
\subfigure[]{\includegraphics [angle=0,width=.3\linewidth] {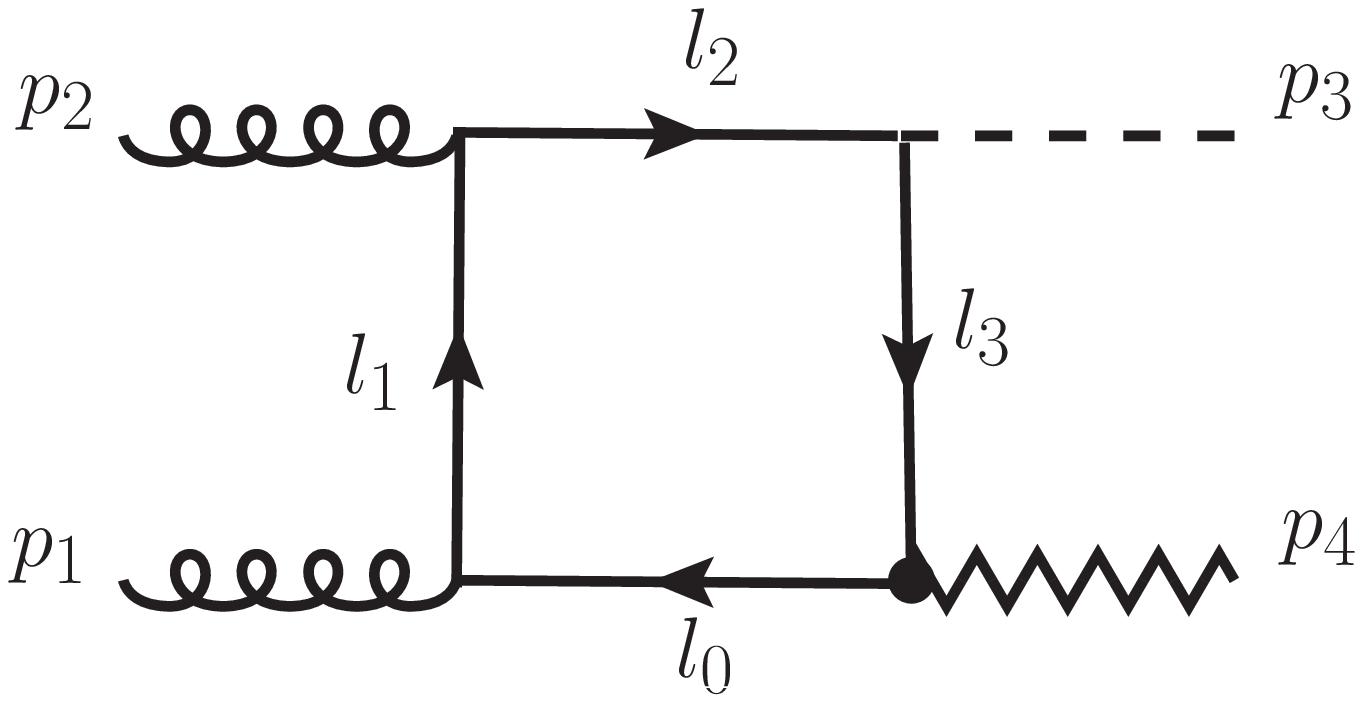}\label{fig:bxa}}
\subfigure[]{\includegraphics [angle=0,width=.3\linewidth] {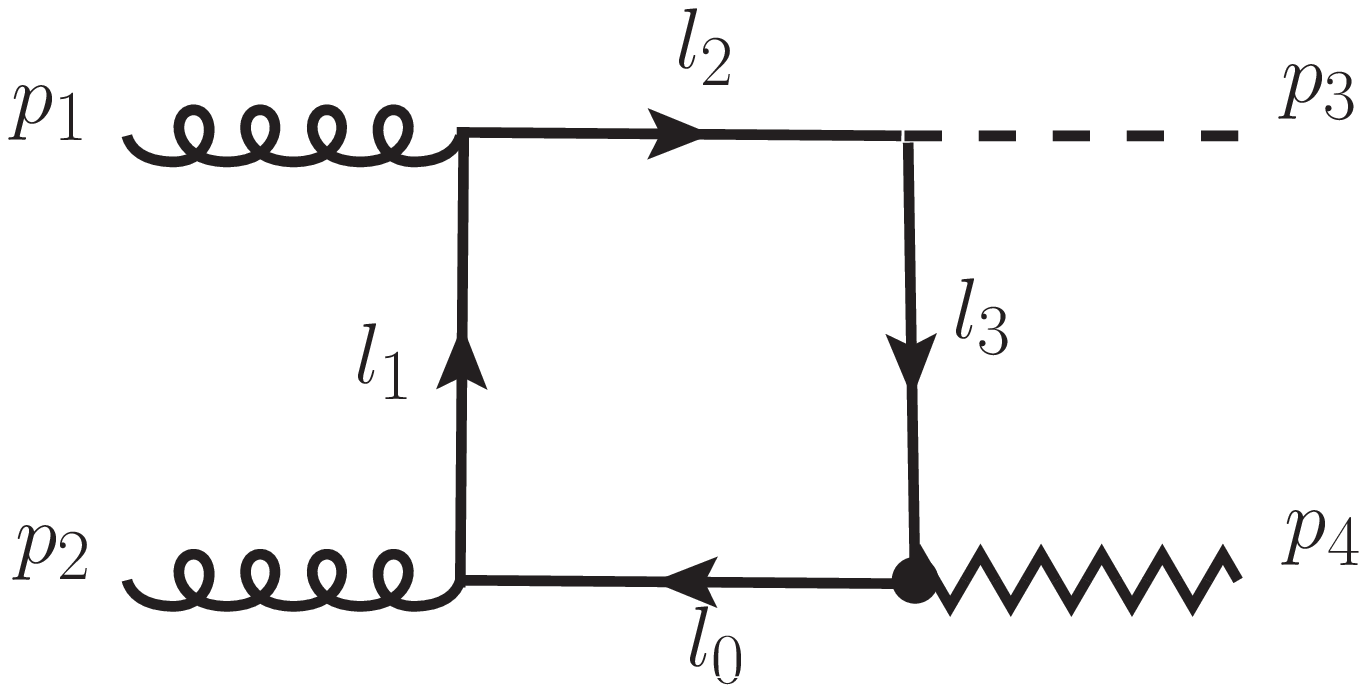}\label{fig:bxb}}
\subfigure[]{\includegraphics [angle=0,width=.3\linewidth] {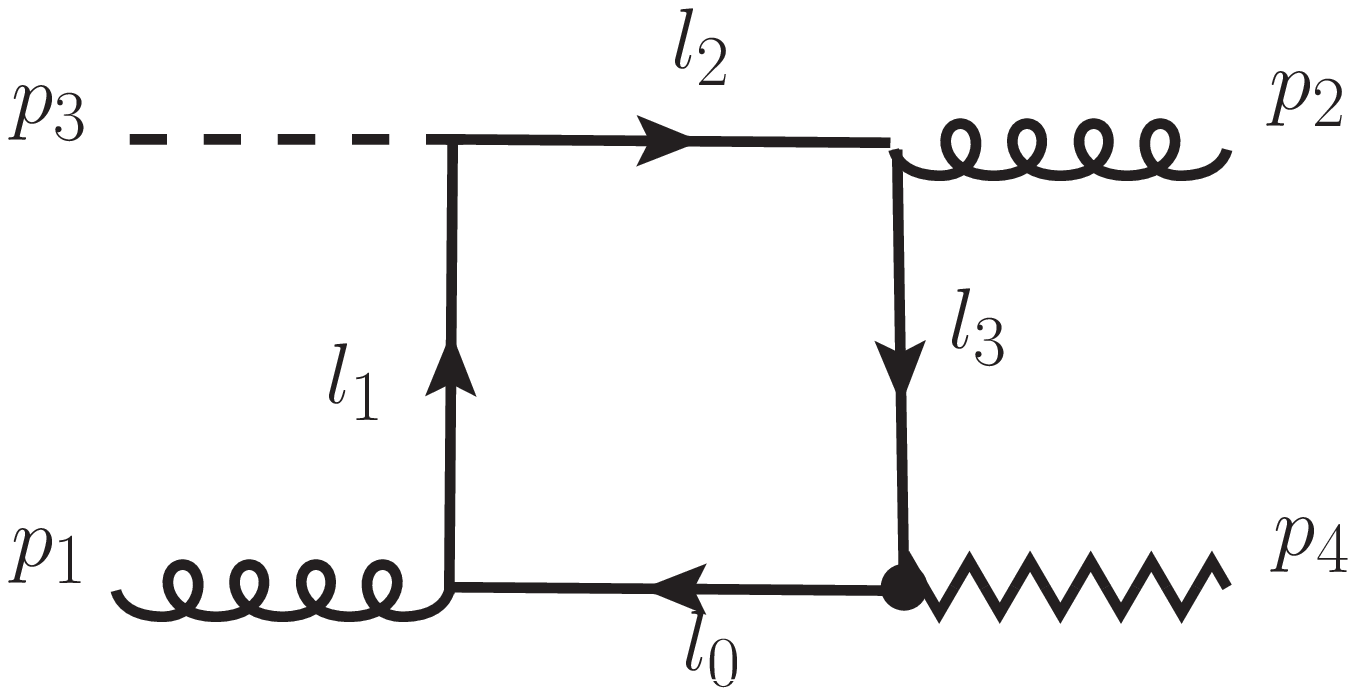}\label{fig:bxc}}
\subfigure[]{\includegraphics [angle=0,width=.3\linewidth] {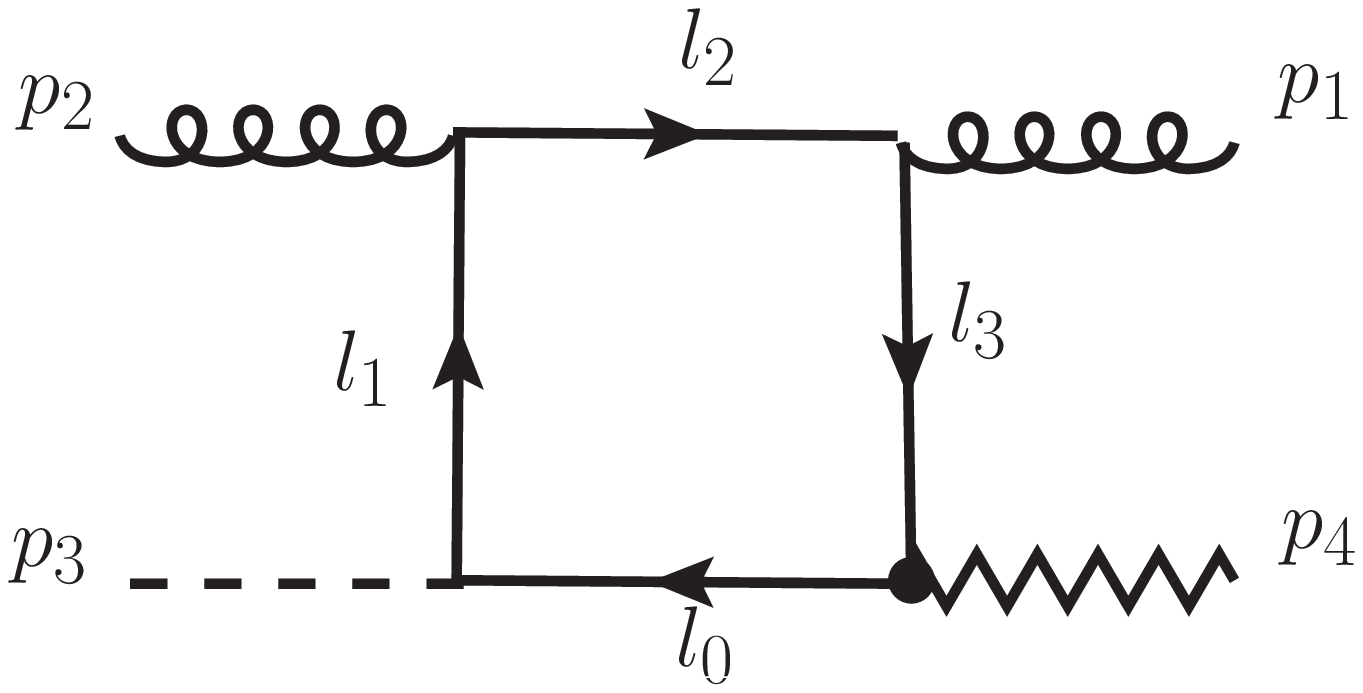}\label{fig:bxd}}
\subfigure[]{\includegraphics [angle=0,width=.3\linewidth] {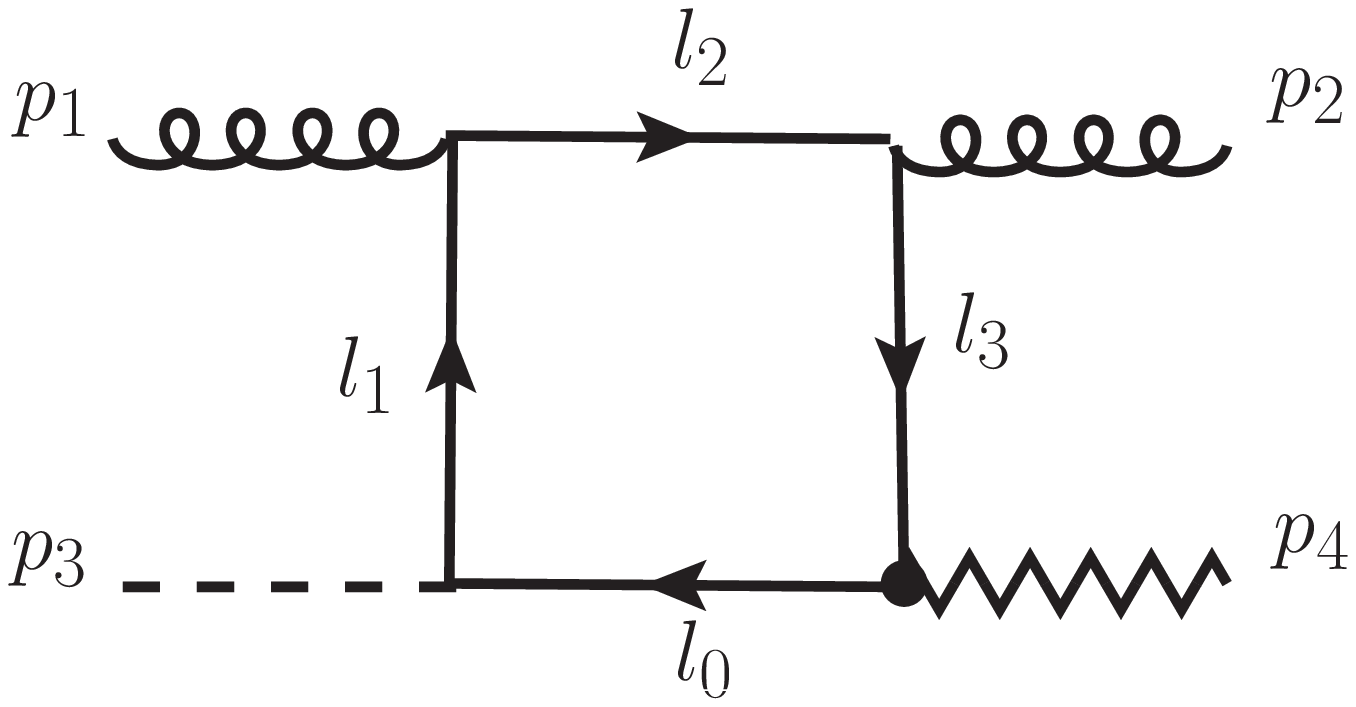}\label{fig:bxe}}
\subfigure[]{\includegraphics [angle=0,width=.3\linewidth] {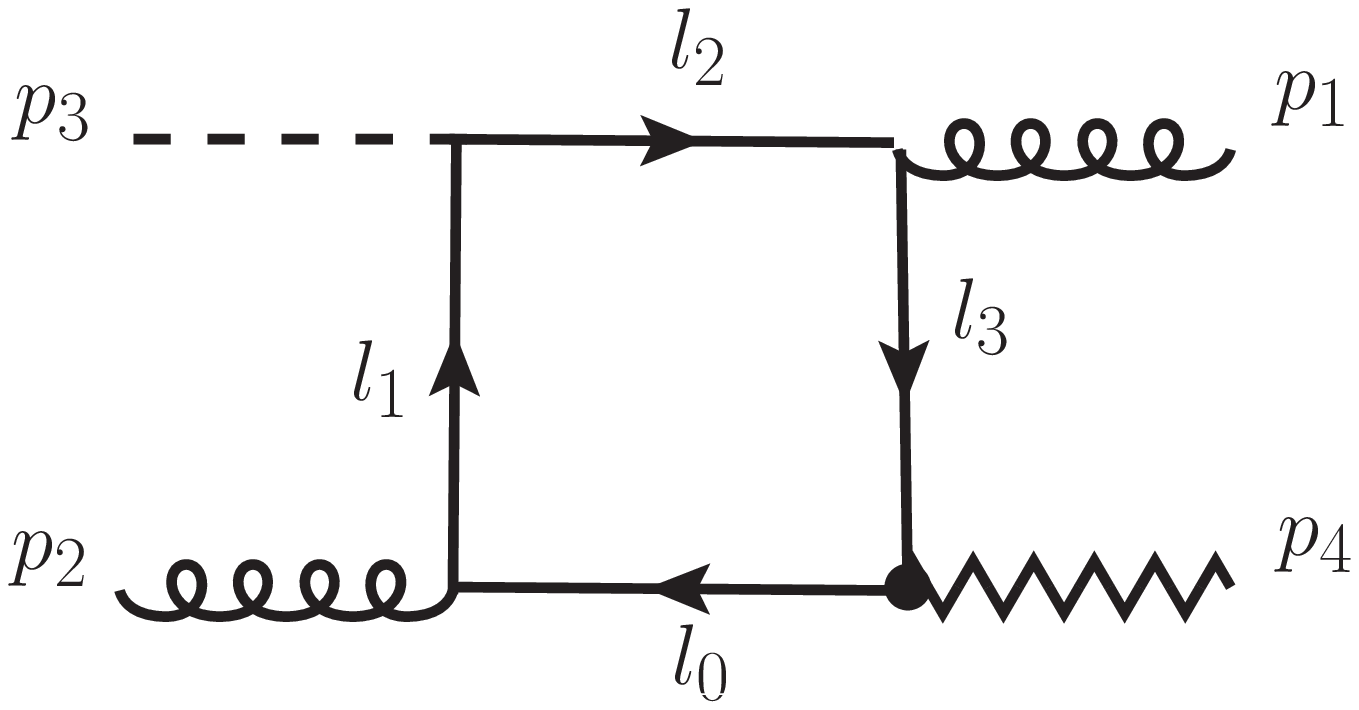}\label{fig:bxf}}
\ec
\caption{Box diagrams that contribute to the process $gg\to h G_{\rm KK}$. The graviton is denoted 
by the zigzag lines. All the external momenta are assumed to be incoming.
}
\label{fig:box}
 \end{figure}

\begin{figure}[!ht]
\bc
\subfigure[]{\includegraphics [angle=0,width=.23\linewidth] {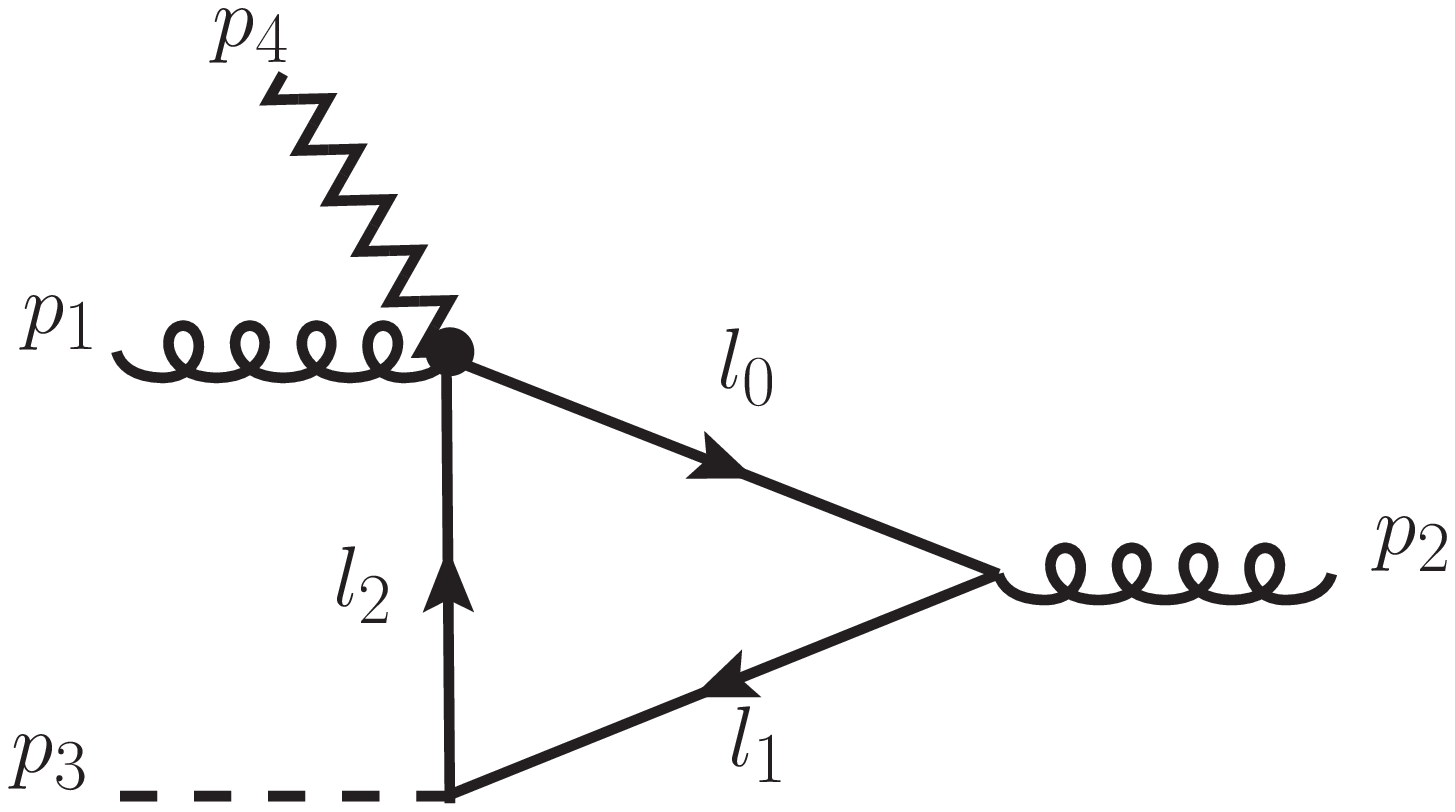}\label{fig:tra}}
\subfigure[]{\includegraphics [angle=0,width=.23\linewidth] {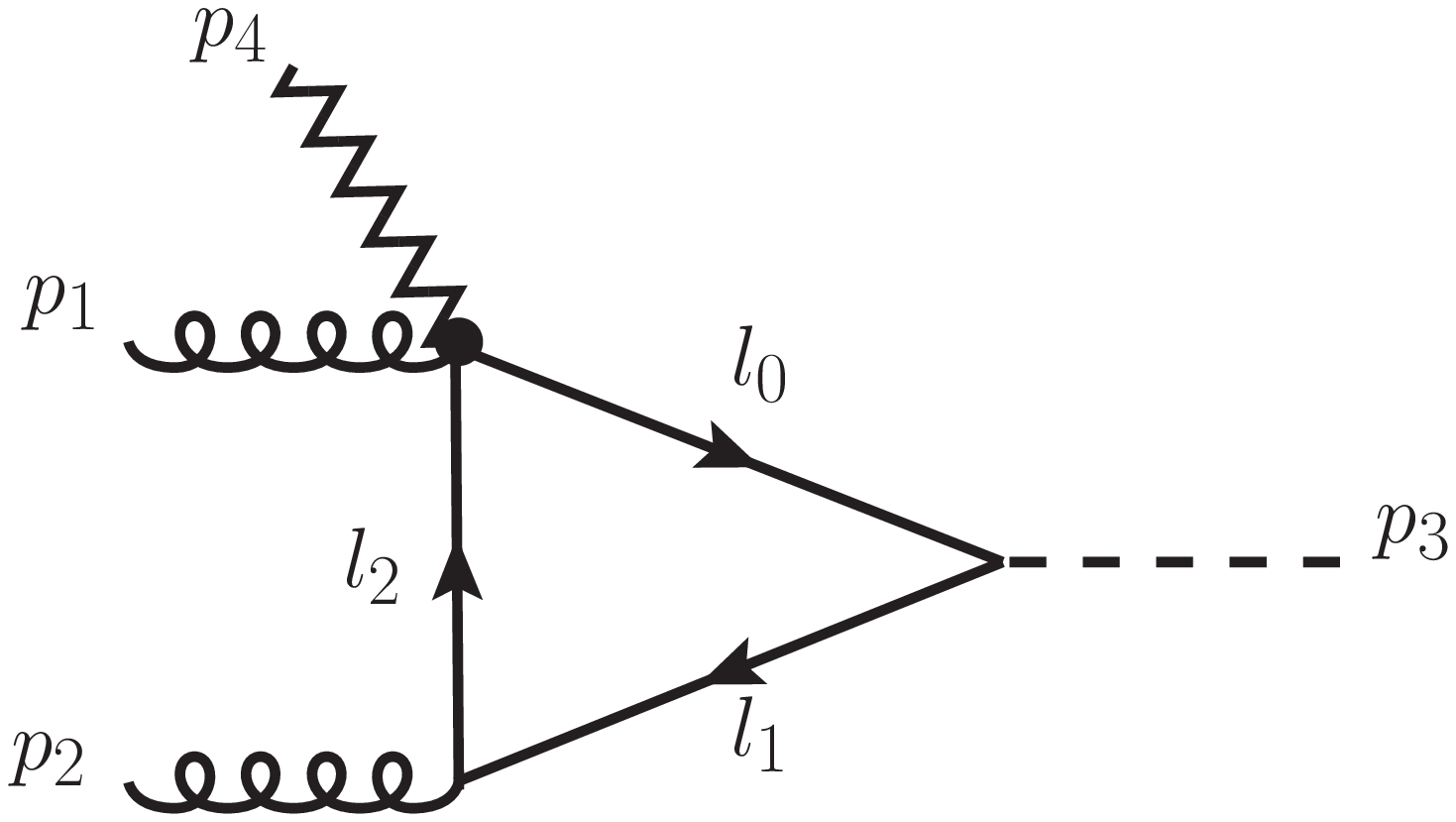}\label{fig:trb}}
\subfigure[]{\includegraphics [angle=0,width=.23\linewidth] {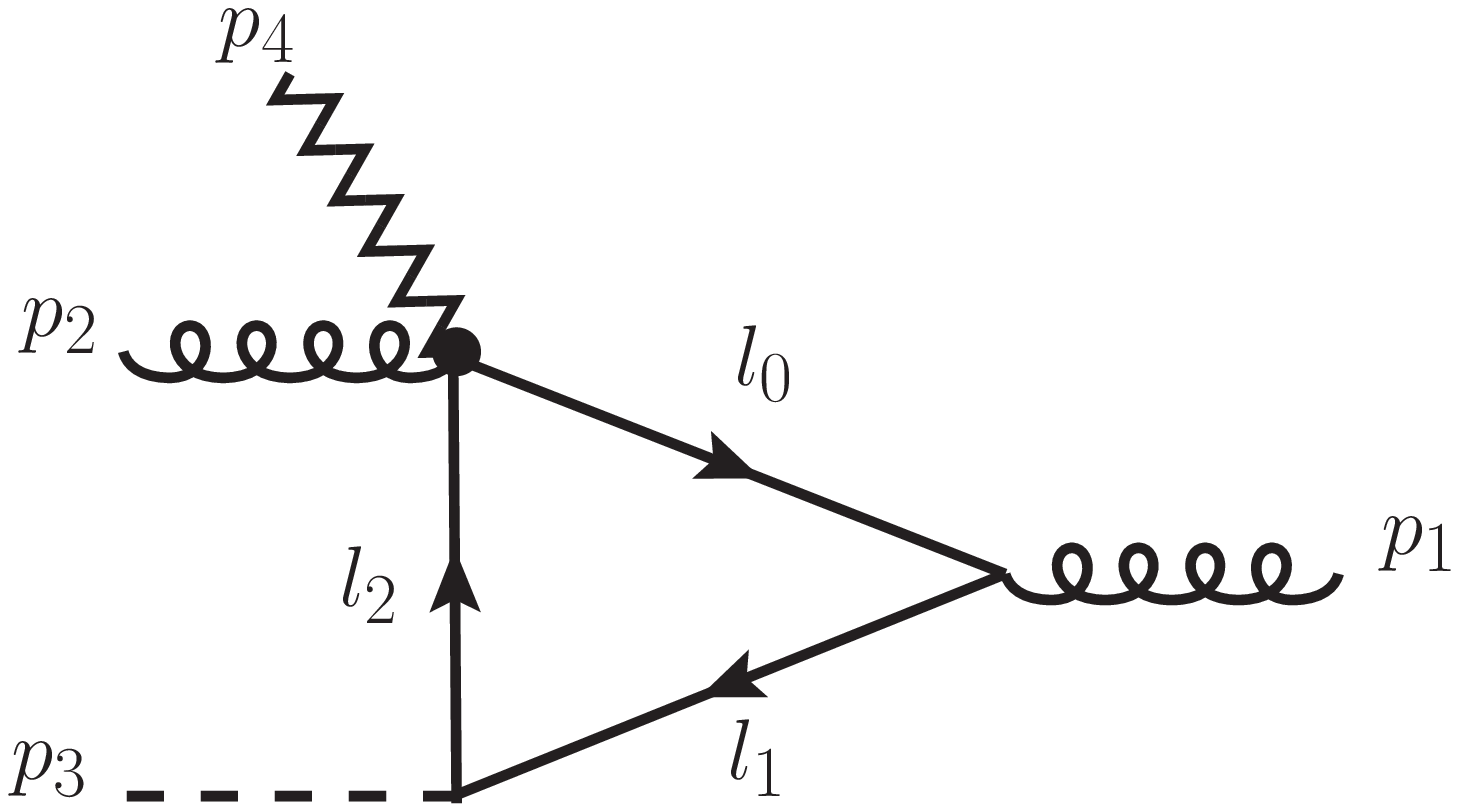}\label{fig:trc}}
\subfigure[]{\includegraphics [angle=0,width=.23\linewidth] {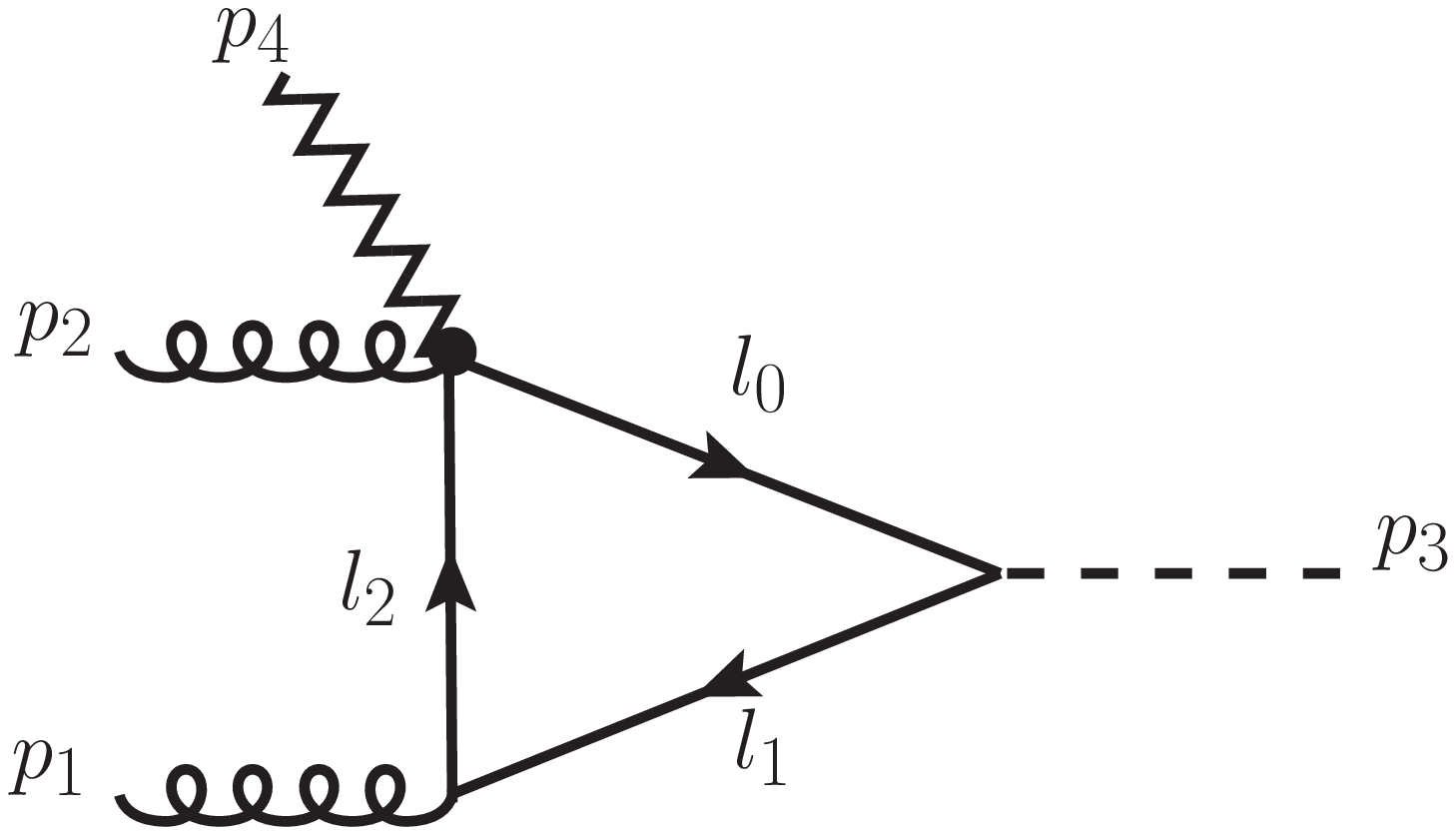}\label{fig:trd}}
\subfigure[]{\includegraphics [angle=0,width=.23\linewidth] {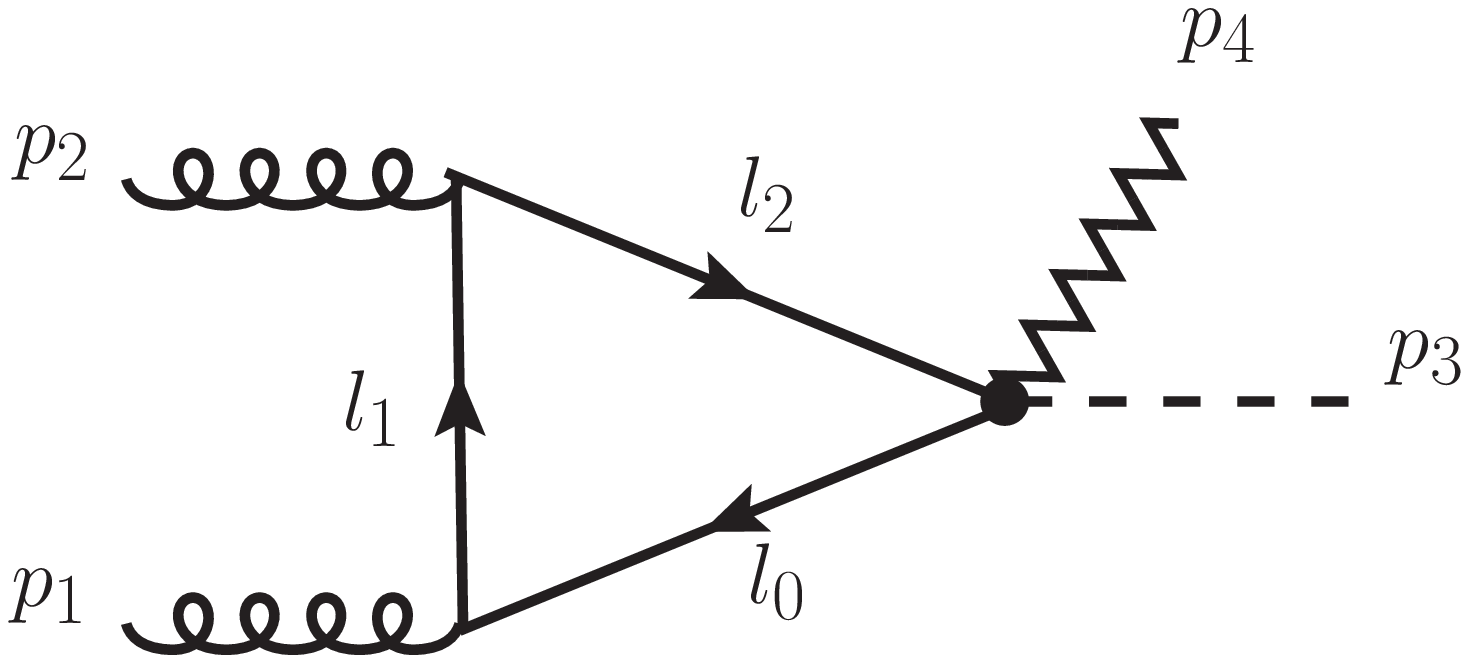}\label{fig:tre}}
\subfigure[]{\includegraphics [angle=0,width=.23\linewidth] {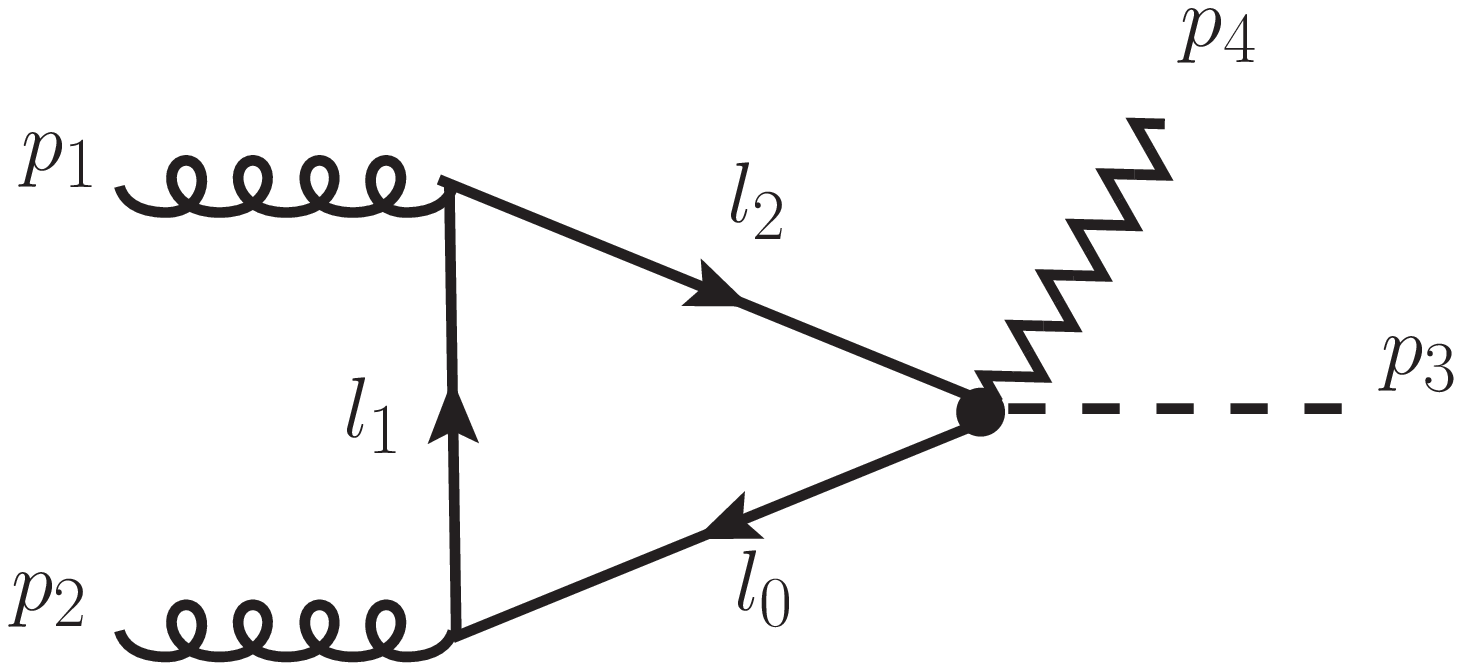}\label{fig:trf}}
\subfigure[]{\includegraphics [angle=0,width=.24\linewidth] {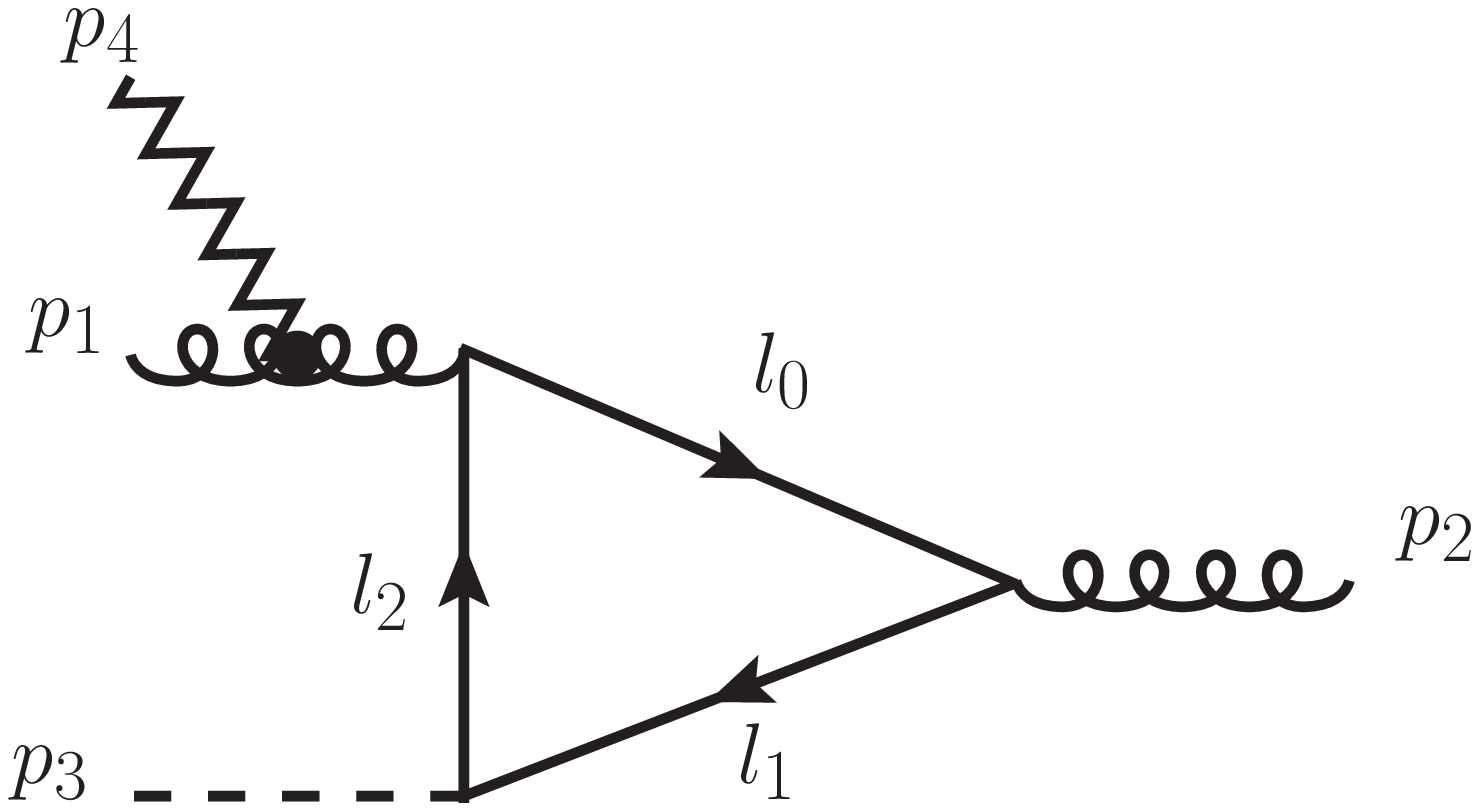}\label{fig:trg}}
\subfigure[]{\includegraphics [angle=0,width=.24\linewidth] {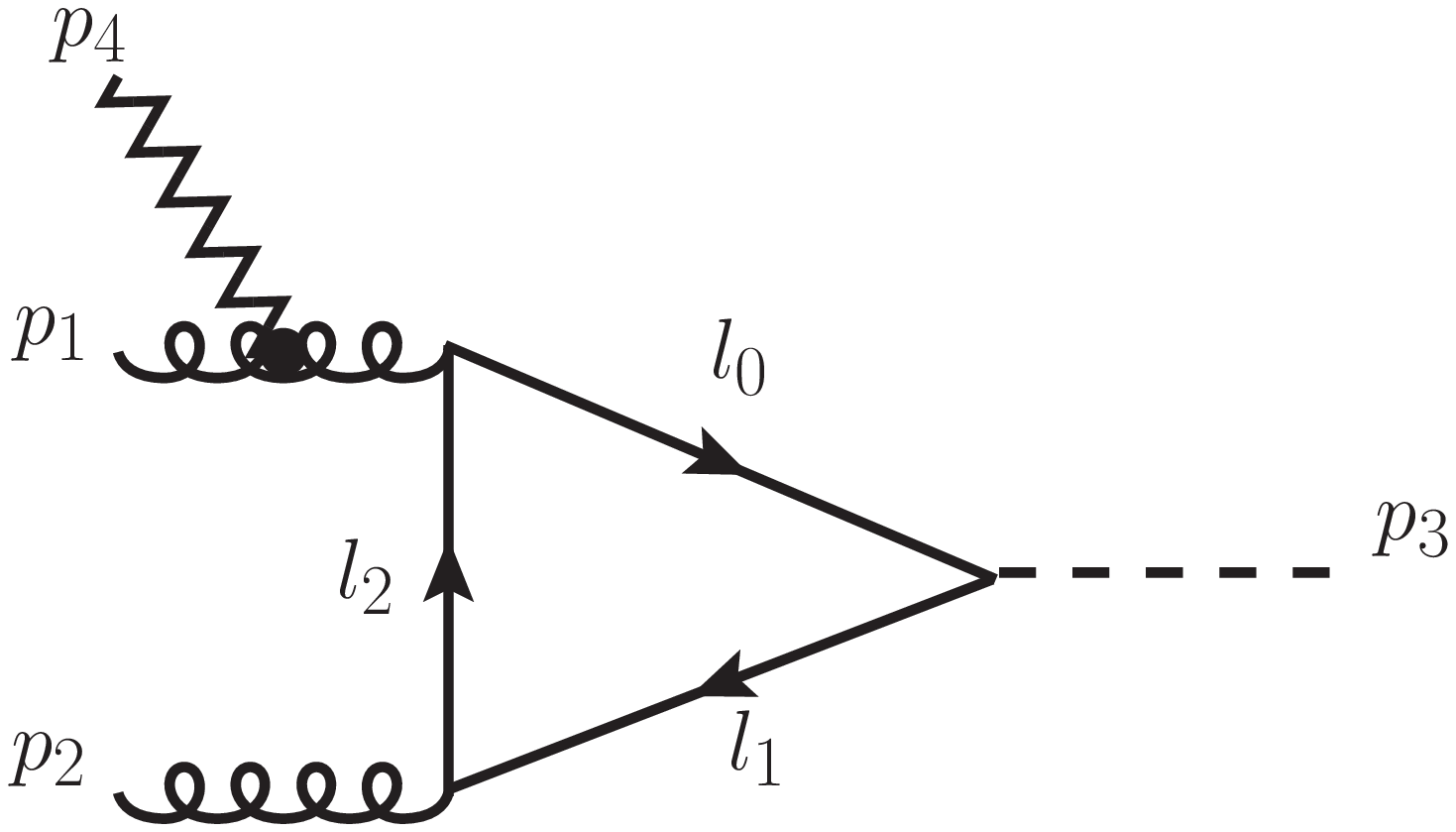}\label{fig:trh}}
\subfigure[]{\includegraphics [angle=0,width=.24\linewidth] {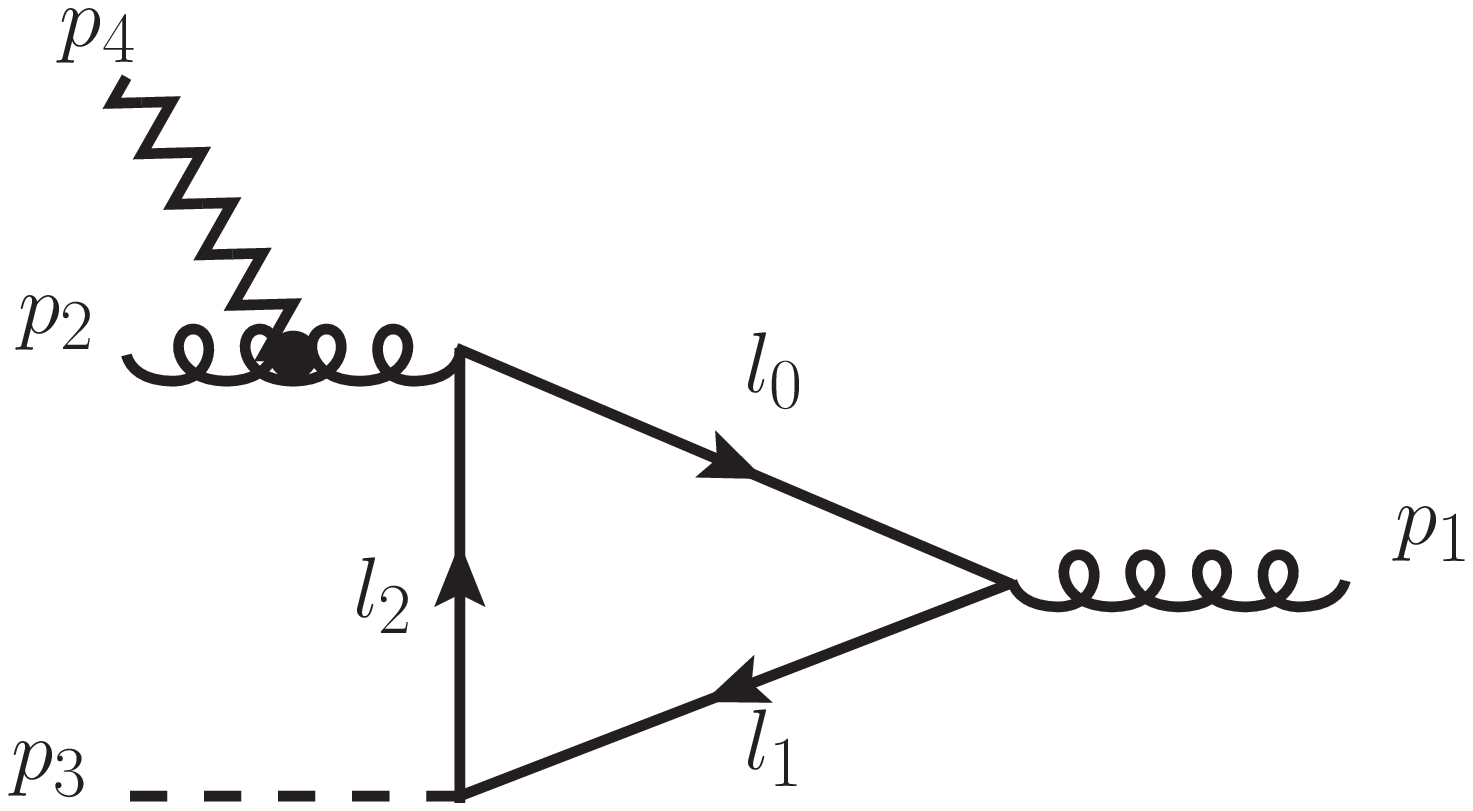}\label{fig:tri}}
\subfigure[]{\includegraphics [angle=0,width=.24\linewidth] {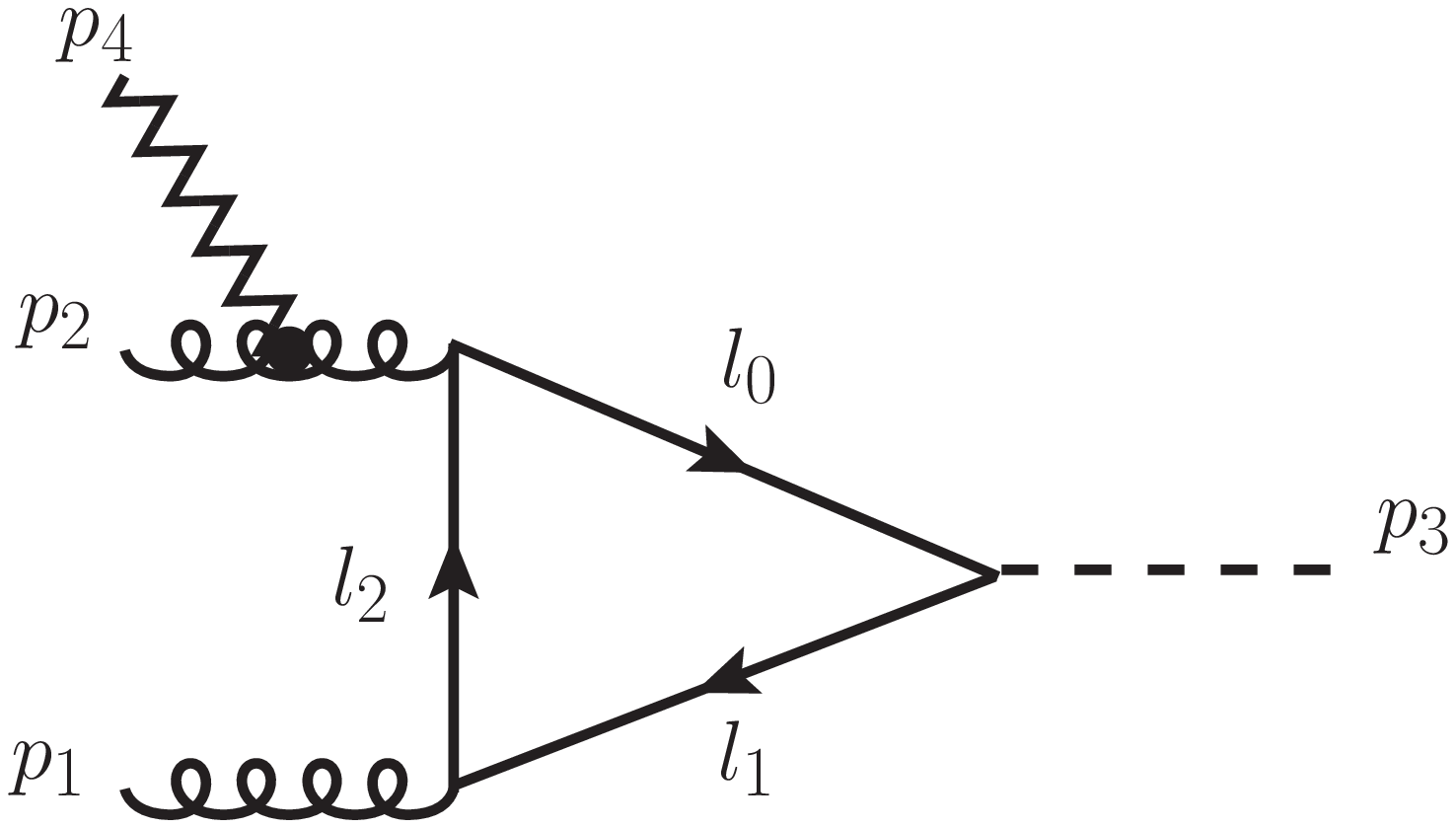}\label{fig:trj}}
\subfigure[]{\includegraphics [angle=0,width=.23\linewidth] {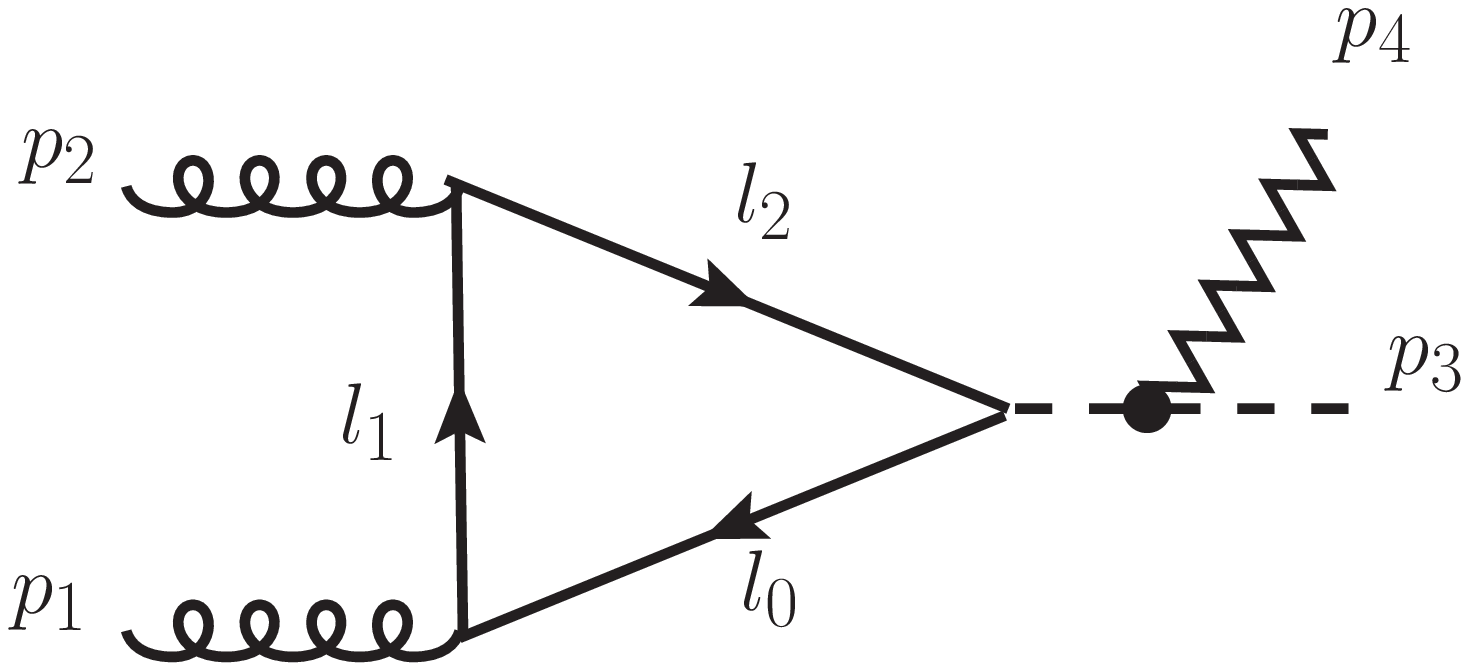}\label{fig:trk}}
\subfigure[]{\includegraphics [angle=0,width=.23\linewidth] {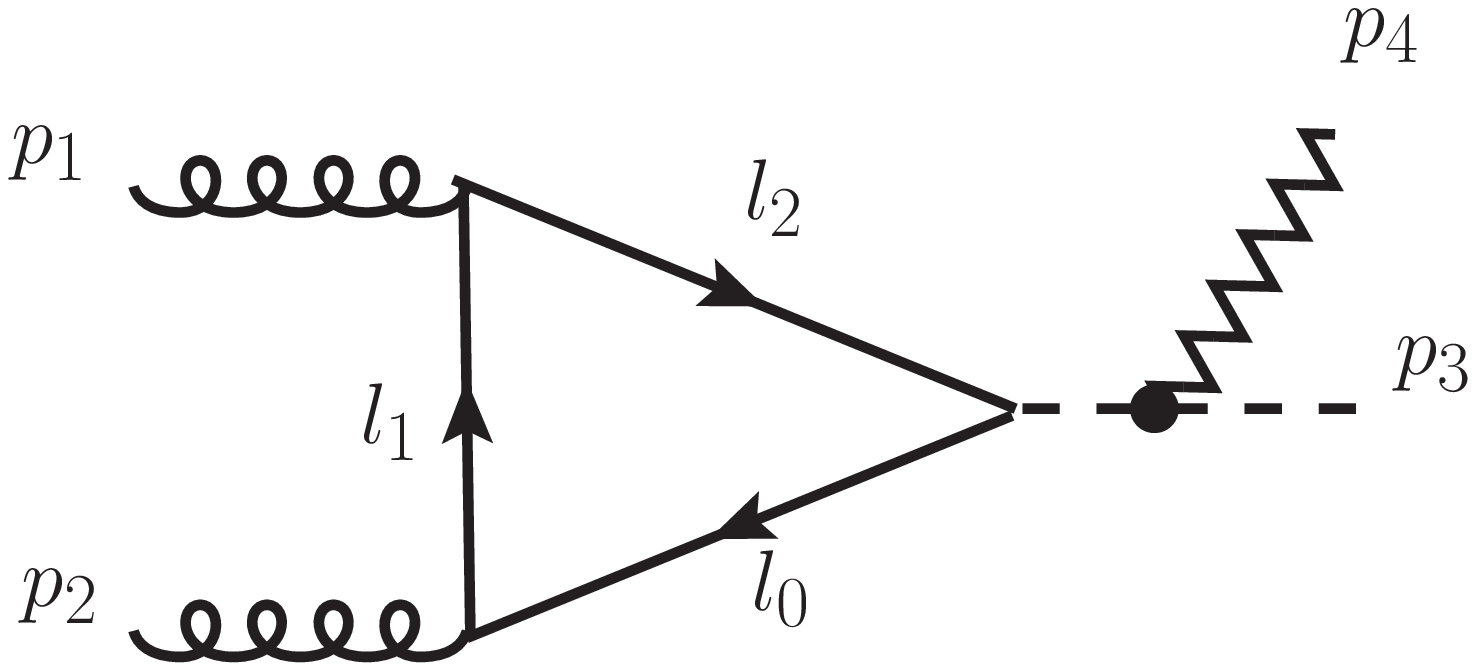}\label{fig:trl}}
\ec
\caption{Triangle diagrams that contribute to the process $gg\to h G_{\rm KK}$. The zigzag lines 
denote the graviton. All the external momenta
are assumed to be incoming. Class I diagrams,
containing a quark-quark-boson-Graviton vertex (either $qqgG_{\rm KK}$ or $qqhG_{\rm KK}$), 
are shown in \subref{fig:tra} - \subref{fig:trf}, and 
Class II diagrams, containing a boson-boson-Graviton vertex 
(either $ggG_{\rm KK}$ or $hhG_{\rm KK}$), are shown in \subref{fig:trg} - \subref{fig:trl}. }
\label{fig:tr}
\end{figure}

Just like the single Higgs production via $gg\to h$ process in the SM, there is no tree level
diagram that contributes to the $gg\to h G_{\rm KK}$ process. The first non vanishing contribution 
comes from the diagrams containing a fermion (quark) loop (at $\mc O(g^2_s \kp y_q)$). However, 
since a Yukawa coupling ($y_q$) is present at the Higgs-quark-quark ($hqq$) vertex, all the 
light-quark loop contributions are negligible. We, therefore, consider the bottom and top quark loop 
contributions only. The relevant Feynman diagrams are shown in Figs. \ref{fig:box} and \ref{fig:tr}. 
Since both the final state particles are 
color singlet, diagrams containing three gluon vertices are absent due to color conservation. 
This leaves us with six different box diagrams (Fig. \ref{fig:box}) and twelve triangle diagrams (Fig. \ref{fig:tr}). However,
using the \emph{charge-conjugation} ($\mc C$) transformation properties, one can see that 
only half of the diagrams are independent. The triangle
diagrams can be grouped into two different classes depending on the graviton vertex present\cite{Han:1998sg}\footnote{
The $hqqG_{\rm KK}$ vertex is not given in Ref. \cite{Han:1998sg}. However, it can be easily derived following the paper
and is given in \ref{sec:newfeynrule}.}
-- i) Class I diagrams contain a quark-quark-boson-Graviton vertex (either $qqgG_{\rm KK}$ or $qqhG_{\rm KK}$), 
 as shown in Figs. \ref{fig:tr}\subref{fig:tra}-\subref{fig:trf}, and ii) Class II diagrams contain a boson-boson-Graviton vertex 
(either $ggG_{\rm KK}$ or $hhG_{\rm KK}$) and are shown in Figs. \ref{fig:tr}\subref{fig:trg}-\subref{fig:trl}.
However, as explained above, the contribution from the triangle diagrams 
with a $hqqG_{\rm KK}$ vertex vanishes.


Feynman rules for the vertices required to calculate these diagrams can be found in Ref. \cite{Han:1998sg}. We display the amplitudes
of two of the box diagrams and three triangle diagrams below. Amplitudes for other diagrams can be generated from these 
prototype diagrams by interchanging appropriate external momenta and polarization vectors. 

\subsection{Prototype Box Diagrams}
The amplitude for the box diagram shown in Fig. \ref{fig:box}\subref{fig:bxa} is given by,
\begin{eqnarray}
\mathcal{M}^{ij}_{box (a)} &=& C^{ij}_{\rm G} \int\frac{d^n l_0}{(2 \pi)^n}\; tr\left[(\slashed l_0+m_t) V_{\rm qqG}^{\al\bt}(l_3,l_0) 
                                 (\slashed l_3+m_t)
                                 (\slashed l_2+m_t) \gamma_{\mu_2}
                                 (\slashed l_1+m_t) \gamma_{\mu_1}\right]\nn\\
                        &&\times\frac{\ve^{\mu_1}(p_1) \ve^{\mu_2}(p_2) \ve^{\rm G}_{\al\bt}(p_4)}{D_0 D_1 D_2 D_3},
\end{eqnarray}
where $i,j$ denote the color indices of the gluons, $\ve$ is the polarization vector of a gluon, 
$\ve^{\rm G}$ is the graviton polarization tensor, $D_i = l_i^2- m_t^2$, $n=4-2\ep$ and
the coupling, 
\ba
C^{ij}_{\rm G} = \frac{1}{8} g_s^2\kp y_t \times \frac{\dl^{ij}}{2}.
\ea
All the vertex factors, $V$'s, are given in \ref{sec:notation}. The amplitude of the diagram given in Fig. \ref{fig:box}\subref{fig:bxb} 
can be obtained from this amplitude by permuting momenta and polarization vectors of the gluons.
Similarly, the amplitude for the box diagram shown in 
Fig. \ref{fig:box}\subref{fig:bxc} is given by,
\begin{eqnarray}
\mathcal{M}^{ij}_{box (c)} &=& C^{ij}_{\rm G} \int\frac{d^n l_0}{(2 \pi)^n}\; tr\left[(\slashed l_0+m_t) V_{\rm qqG}^{\al\bt}(l_3,l_0) 
                                 (\slashed l_3+m_t) \gamma_{\mu_2}
                                 (\slashed l_2+m_t) 
                                 (\slashed l_1+m_t) \gamma_{\mu_1}\right]\nn\\
                        &&\times\frac{\ve^{\mu_1}(p_1) \ve^{\mu_2}(p_2) \ve^{\rm G}_{\al\bt}(p_4)}{D_0 D_1 D_2 D_3}.
\end{eqnarray}
The other three box diagrams give contributions identical to the first three diagrams. 

\subsection{Prototype Triangle Diagrams}
The amplitude for the triangle diagrams shown in Figs. \ref{fig:tr}\subref{fig:tra}, \ref{fig:tr}\subref{fig:trg} and \ref{fig:tr}\subref{fig:trk} in the 
Feynman gauge are given as,
\begin{eqnarray}
\mathcal{M}^{ij}_{tri (a)} &=& -2C^{ij}_{\rm G} \int\frac{d^n l_0}{(2 \pi)^n}\; tr\left[(\slashed l_0+m_t) V_{\rm qqgG}^{\al\bt;\mu_1}
                                 (\slashed l_2+m_t) 
                                 (\slashed l_1+m_t) \gamma^{\mu_2}\right]\nn\\
                           &&\times\frac{\ve_{\mu_1}(p_1) \ve_{\mu_2}(p_2) \ve^{\rm G}_{\al\bt}(p_4)}{D_0 D_1 D_2},\\
\mathcal{M}^{ij}_{tri (g)} &=& -4C^{ij}_{\rm G} \int\frac{d^n l_0}{(2 \pi)^n}\; tr\left[(\slashed l_0+m_t) \gamma_{\rho}
                                 (\slashed l_2+m_t) 
                                 (\slashed l_1+m_t) \gamma^{\mu_2}\right]  \nonumber \\
                        &&\times \frac{V_{\rm ggG}^{\al\bt; \mu_1 \rho}(p_1,p_2+p_3)}{(p_2+p_3)^2}
                          \times\frac{\ve_{\mu_1}(p_1) \ve_{\mu_2}(p_2) \ve^{\rm G}_{\al\bt}(p_4)}{D_0 D_1 D_2},\\
\mathcal{M}^{ij}_{tri (k)} &=& -4C^{ij}_{\rm G} \int\frac{d^n l_0}{(2 \pi)^n}\; tr\left[(\slashed l_0+m_t) 
                                 (\slashed l_2+m_t) \gamma^{\mu_2}
                                 (\slashed l_1+m_t) \gamma^{\mu_1}\right]\nn \\
                        &&\times \frac{V_{\rm hhG}^{\al\bt}(p_1+p_2,p_3)}{(p_1+p_2)^2-m_h^2}
                       \times\frac{\ve_{\mu_1}(p_1) \ve_{\mu_2}(p_2) \ve^{\rm G}_{\al\bt}(p_4)}{D_0 D_1 D_2}.
\end{eqnarray}
The contribution of all other triangle diagrams can be obtained by appropriate permutations of the momenta and polarization vectors.

To compute these amplitudes, we first compute the traces associated with the top quark loop by using symbolic manipulation
program, FORM \cite{Vermaseren:2000nd}. At this stage, the amplitude contains tensor integrals,
4-rank tensor-box integral ($D^{\mu \nu \rho \sigma}$) being the most complicated one, 
\begin{equation}
D^{\mu \nu \rho \sigma} = \int \frac{d^n l_0}{(2 \pi)^n} \frac{l_0^{\mu} l_0^{\nu} l_0^{\rho} l_0^{\sigma}}{D_0 D_1 D_2 D_3}.
\end{equation}
We reduce the tensor integrals into the standard scalar integrals -- $A_0$, $B_0$, $C_0$ and $D_0$ \cite{'tHooft:1978xw} 
 using the reduction scheme 
developed by Oldenborgh and Vermaseren \cite{vanOldenborgh:1989wn}. The algorithm described in Ref. \cite{vanOldenborgh:1989wn} 
was first coded using Mathematica. Then after the reduction, appropriate Fortran routines were obtained \cite{agrawal}.
After the full reduction the amplitude has the following generic structure,
\begin{equation}
 \mathcal{M} = \sum_i \left(d_i D_0^i\right) +  \sum_i \left(c_i C_0^i\right) +  \sum_i \left(b_i B_0^i\right) +  \sum_i \left(a_i A_0^i\right)  + \mathcal{R},  
\end{equation}
where $\mathcal{R}$ is the rational term coming from the UV regularization of tensor integrals and the index $i$ stands for different momentum combinations
that enter in these scalars.
All the needed scalar integrals (with massive internal lines) are called from FF library \cite{vanOldenborgh:1990yc}. The amplitude is thus a function of
external momenta and polarizations. We consider the helicity basis for the polarization vectors to calculate the amplitude. We explicitly 
check the symmetries of the helicity amplitudes to ensure the correctness of our calculation.

With the above amplitude, we perform integrations over 
the two body phase space, momentum fractions ($x_{1}/x_2$) of the initial state gluons and
over the graviton mass parameter in the continuum approximation (for the ADD model).
As a cautionary check, we have performed a few  tests with our program. The amplitude should
be free of divergences and also be gauge invariant. Because of the massive top quark in the
loop, there are no infrared divergences. We have checked the UV finiteness and gauge invariance.
\begin{enumerate}
\item {\it UV Finiteness:}
Individual box and triangle diagrams can be UV divergent, but the total amplitude should be UV
finite.
We have tested the UV finiteness of the total amplitude by varying the renormalization scale ($\m_{\rm R}$) 
over ten orders of magnitude. We find that the amplitude is independent of the 
actual value of $\m_{\rm R}$.  The triangle and box amplitudes are separately UV finite. In fact, each triangle
diagram is UV finite by itself.  This behavior is similar to that of the $gg\to h$ amplitude which is UV finite.    

\item {\it Gauge Invariance:}
We have checked the gauge invariance of the amplitude with respect to both the gluons. This has been done by replacing the polarization 
vector of either of the gluons by its momentum ($\ve^\m(p_i) \to p^\m_i$). This, as expected, makes the amplitude
zero. We observe that some of the triangle diagrams are separately gauge invariant with respect to both
the gluons. To ensure the correctness of their contribution toward the full amplitude, 
we perform gauge invariance check with respect to the graviton polarization also.

\end{enumerate}

\subsection{Calculation in the Effective Theory}\label{subsec:heft}
\begin{figure}[!t]
\bc
\subfigure[]{\includegraphics [angle=0,width=.24\linewidth] {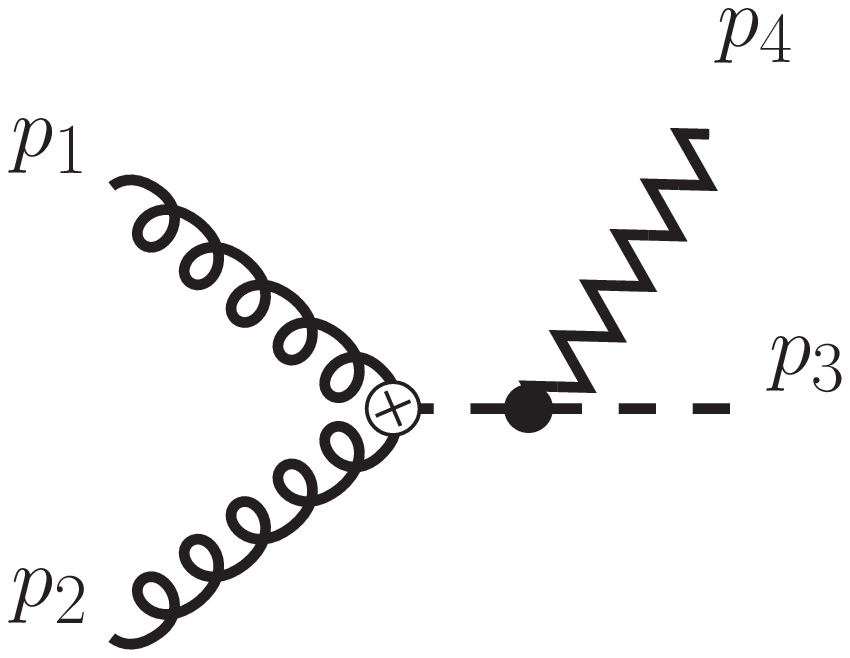}\label{fig:hefta}}
\subfigure[]{\includegraphics [angle=0,width=.24\linewidth] {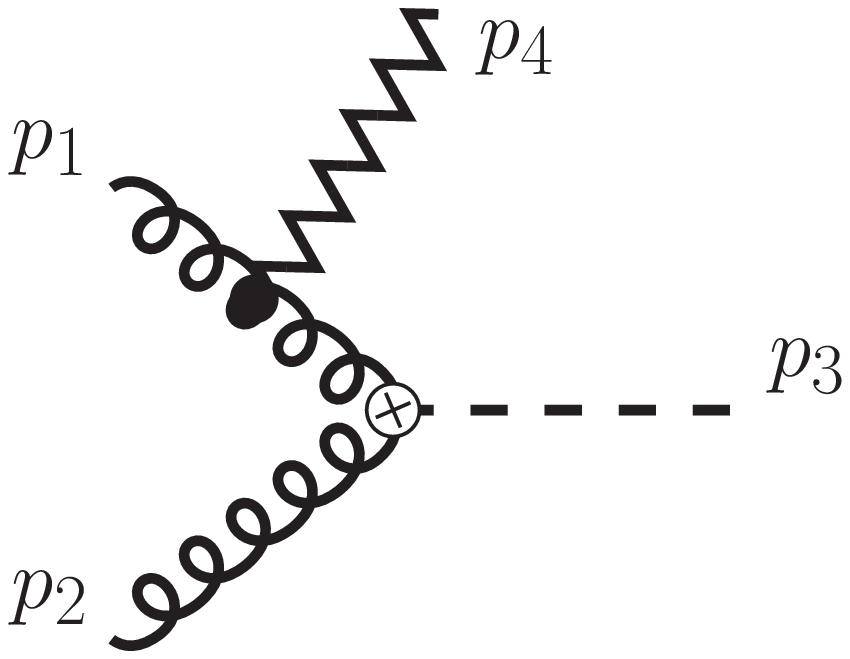}\label{fig:heftb}}
\subfigure[]{\includegraphics [angle=0,width=.24\linewidth] {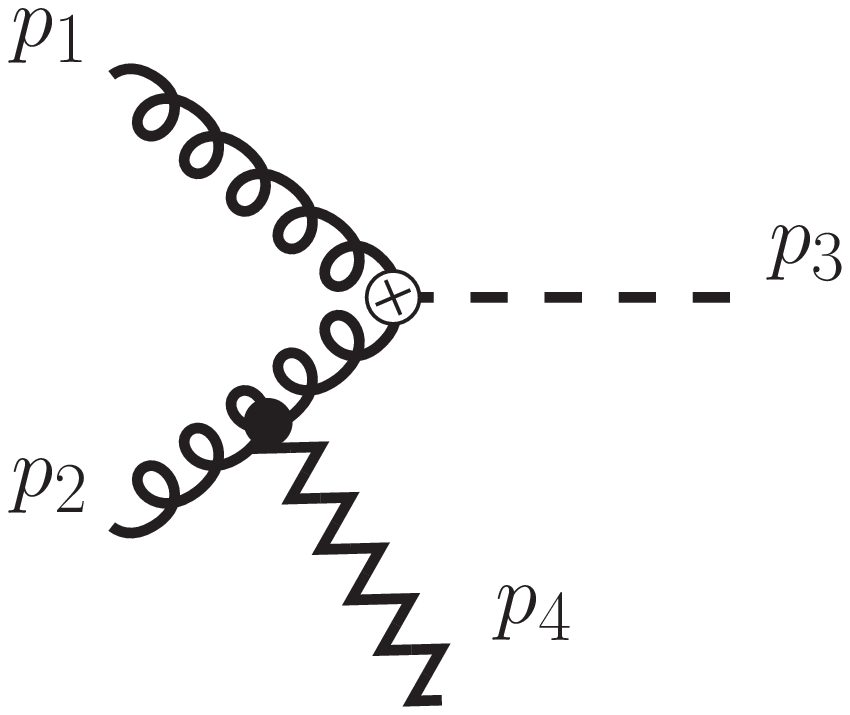}\label{fig:heftc}}
\subfigure[]{\includegraphics [angle=0,width=.24\linewidth] {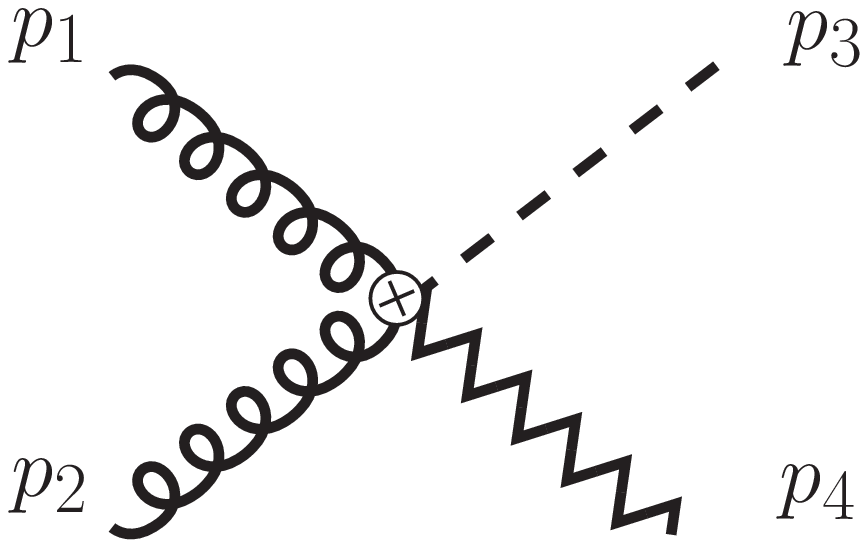}\label{fig:heftd}}
\ec
\caption{Diagrams that contribute to the process $gg\to h G_{\rm KK}$ in the effective theory.}
\label{fig:heftdiags}
 \end{figure}

It is well known that in the SM it is possible to integrate out the top quark loop contribution to the $gg\to h$ process in the 
heavy quark limit, {\it i.e.,} $m_t \gg m_h/2$ and describe the Higgs-gluon interaction by an effective Lagrangian \cite{Rizzo:1979mf},
\ba
\mc L_{\rm eff} = -\frac14 g_h\, G^a_{\m\n}G^{a\,\m\n} h\,, \label{eq:heft}
\ea
where $G^a_{\m\n}$ is the gluon field-strength tensor. The effective coupling, $g_h$ is given as,
\ba
g_h = \frac{\al_s}{3\pi v} \left[1 + \mc O(\tau_h)
\right]\,,
\ea 
where $\tau_h = m^2_h/4m_t^2$ and $v$ is the vacuum expectation value of the Higgs boson field. 
In principle, one can compute the $gg\to hG_{\rm KK}$ process also using this effective theory. 
Within this theory the diagrams that contribute are shown in Fig. \ref{fig:heftdiags}.
The vertex factor for the diagram in Fig. \ref{fig:heftd} can be obtained by computing the 
contribution of the Lagrangian given in Eq. \ref{eq:heft}
to the stress-energy tensor. It is given in \ref{sec:heftfeynrule}. 
 
\section{Results}\label{sec:results}

In this section we present the numerical results for the LHC with $\sqrt{s} = 14$ TeV.
 The ADD model has two parameters -- the cut-off
 scale, $M_{\rm S}$, and the number of extra dimensions, $d$.
We find the dependence of the cross section on these parameters. In addition to these parameters, the cross section 
also depends on the choice of the parton distribution functions and the choice of renormalization/factorization
scale. We choose the LO CTEQ6L1 PDFs\cite{Pumplin:2002vw}. 
For the factorization/renormalization scale, we choose the transverse mass of the Higgs boson,  $\sqrt{m_h^2 + \left(p^h_{\rm T}\right)^2}$. 
Later we comment on the dependence of the cross section on the choice of PDF's or the factorization/renormalization scale.

To compute the cross section, we apply the following cuts on the transverse momentum and rapidity of the Higgs boson: 
$p_{\rm T}^h > 20$ GeV, $|\eta^h| < 2.5$. In case of the ADD model, we apply one extra cut on the invariant mass of the 
outgoing particles: $M(hG_{\rm KK}) \leq M_{\rm S}$. This is the truncated scheme\cite{Giudice:1998ck}. Later, we also comment
on the untruncated scheme.

 \begin{figure}[!ht]
\bc
\includegraphics [angle=0,width=.65\linewidth] {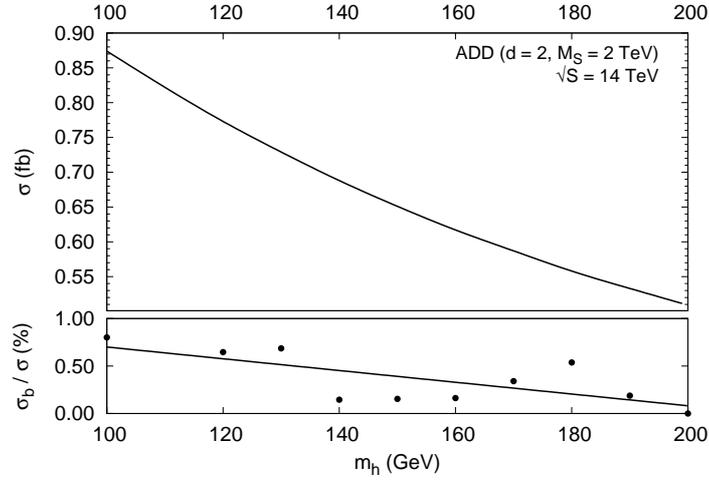}
\ec
\caption{Variation of the cross section ($\s$) with the mass of the Higgs boson ($m_h$) in the ADD model. Here the cutoff scale, $M_{\rm S} = 2$ TeV
and number of extra dimensions, $d = 2$. 
In the lower part of the plot we show the $b$-loop contribution ($\s_b$) to $\s$ as percentage of it. 
The line is a linear fit for the points.}
\label{fig:mh-X}
\end{figure}
\begin{figure}[!ht]
\bc
\includegraphics [angle=0,width=.65\linewidth] {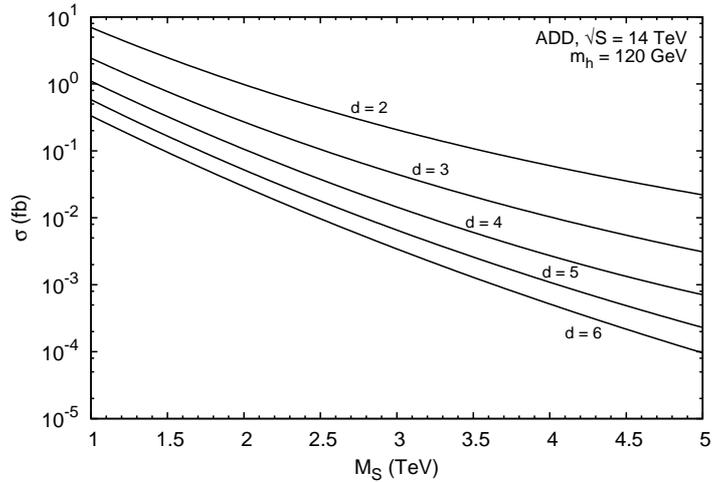}
\ec
\caption{Variation of the cross section ($\s$) with the cutoff scale ($M_{\rm S}$) of the ADD model for different numbers of extra 
dimensions, $d$. For this plot the Higgs mass, $m_h = 120$ GeV.}
\label{fig:ms-X}
\end{figure}
\begin{figure}[!ht]
\bc
\includegraphics [angle=0,width=.65\linewidth] {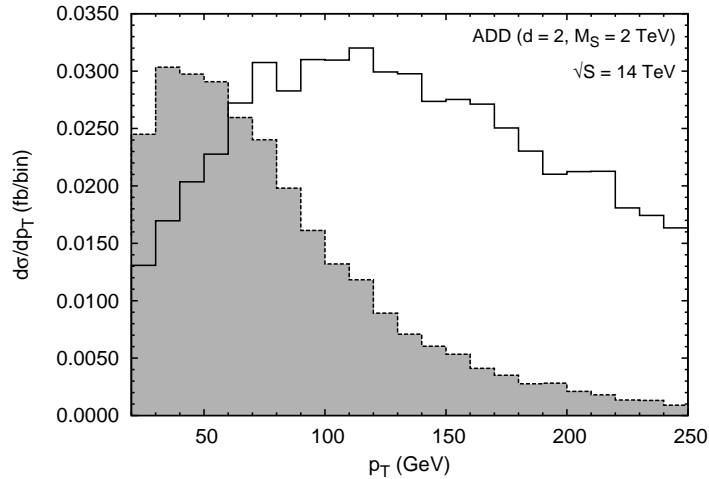}
\ec
\caption{Distribution of the transverse momentum of the Higgs boson in the ADD model for $d=2$ and  $M_{\rm S}$ = 2 TeV. 
Here the Higgs mass, $m_h = 120$ GeV. The shaded curve is the $p_{\rm T}$ distribution
of the Higgs for the SM background $pp\to hZ$ with Z decaying into pair of neutrinos obtained using Madgraph5 \cite{Alwall:2011uj}. 
The background curve is scaled by a factor of 500.}
\label{fig:pt-X}
\end{figure}
\begin{figure}[!ht]
\bc
\subfigure[]{\includegraphics [angle=0,width=.49\linewidth] {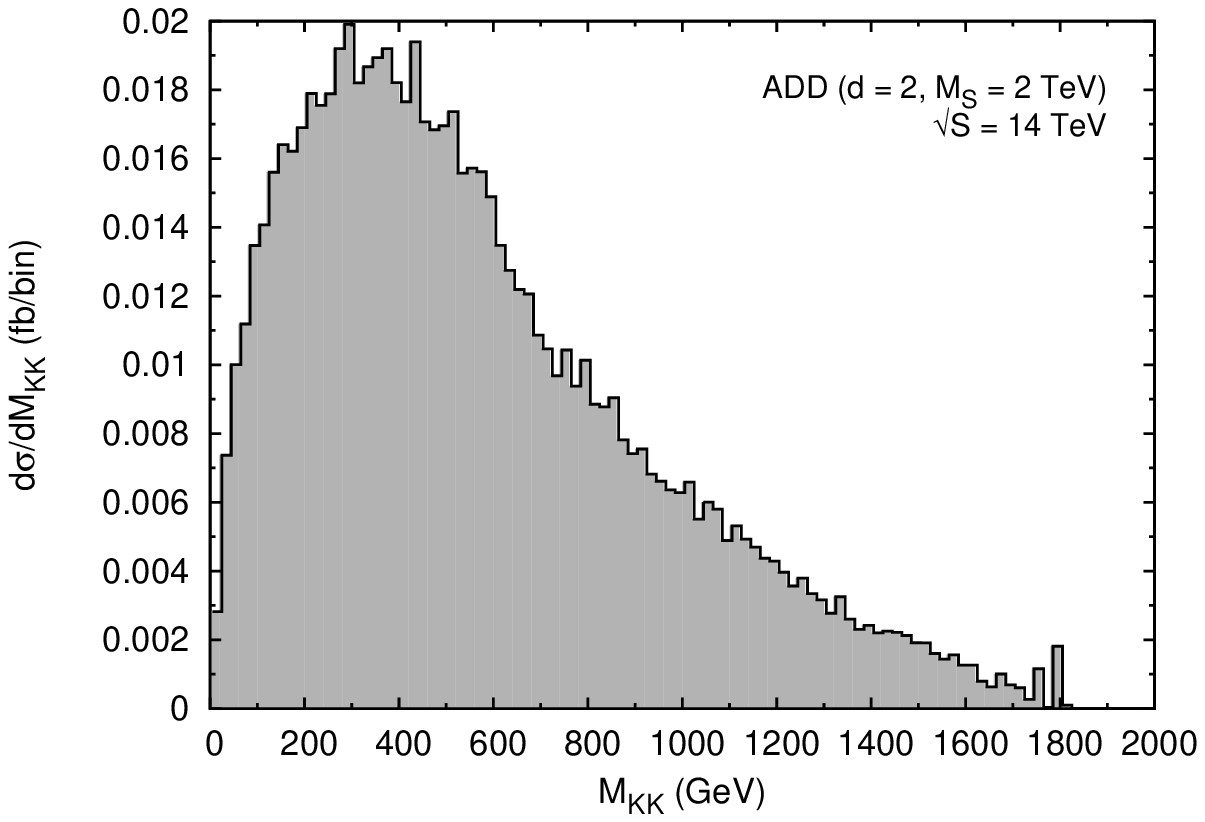}\label{fig:mkk-Xa}}
\subfigure[]{\includegraphics [angle=0,width=.49\linewidth] {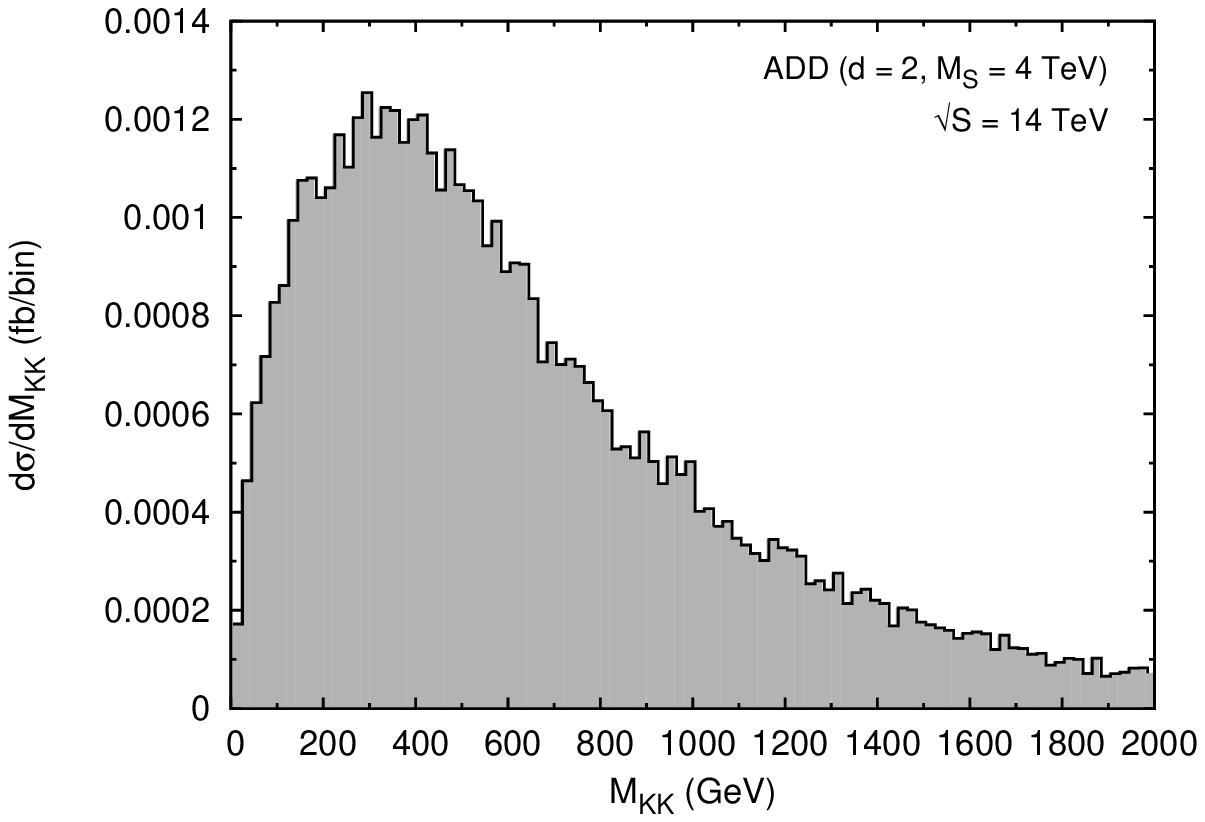}\label{fig:mkk-Xb}}\\
\subfigure[]{\includegraphics [angle=0,width=.49\linewidth] {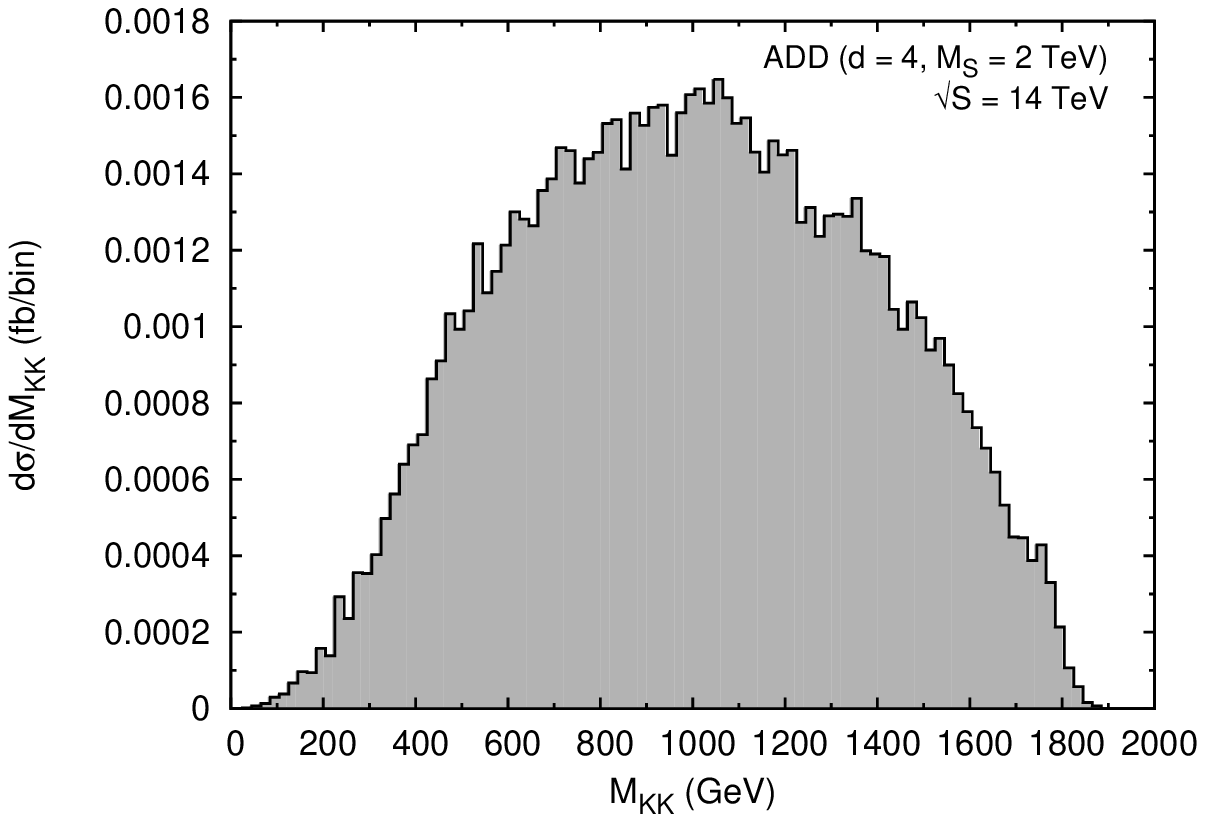}\label{fig:mkk-Xc}}
\subfigure[]{\includegraphics [angle=0,width=.49\linewidth] {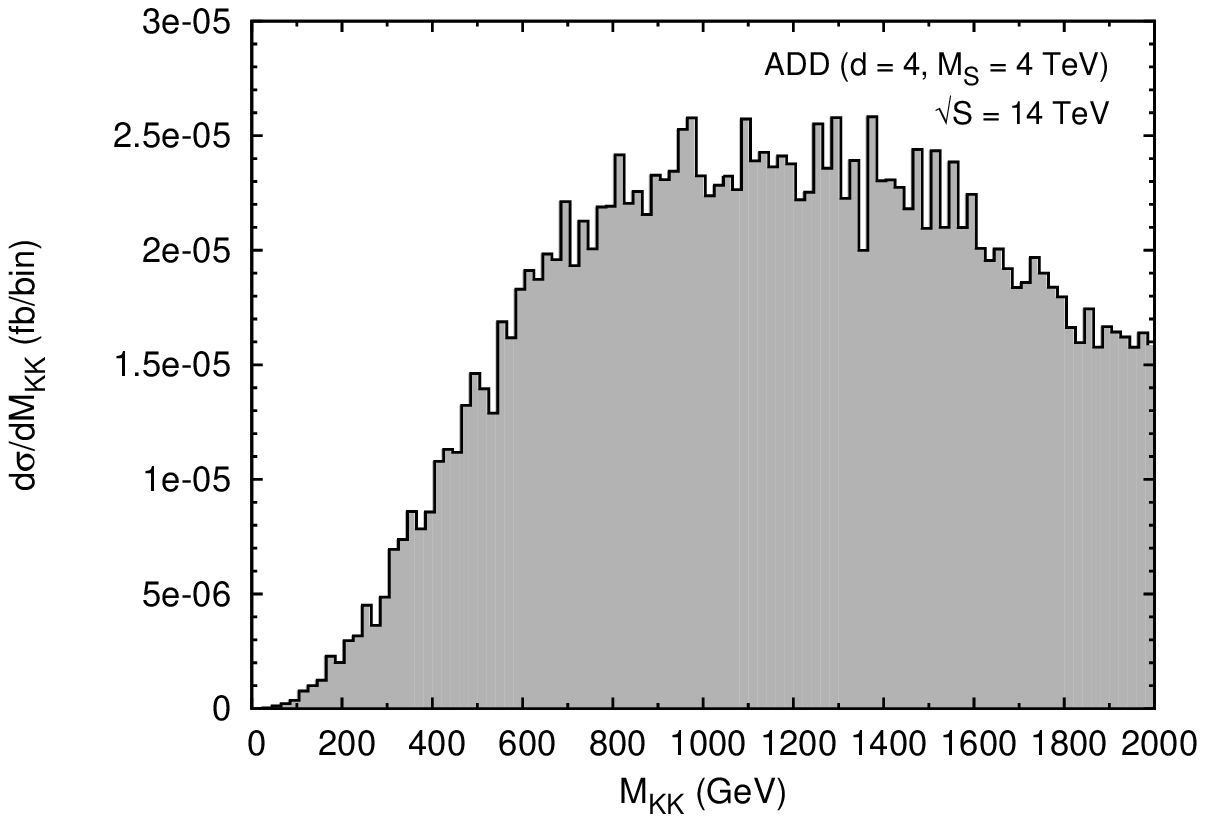}\label{fig:mkk-Xd}}
\caption{Relative contributions of different KK modes in the ADD model for different 
cutoff scales and numbers of extra dimensions -- (a)  $d = 2$, $M_{\rm S}$ = 2 TeV, 
(b) $d = 2$, $M_{\rm S}$ = 4 TeV,  (c) $d = 4$, $M_{\rm S}$ = 2 TeV  and (d) 
 $d = 4$, $M_{\rm S}$ = 4 TeV. For these plots the Higgs mass, $m_h = 120$ GeV.}
\label{fig:mkk-X}\ec \end{figure}
\begin{figure}[!ht]
\bc
\includegraphics [angle=0,width=.65\linewidth] {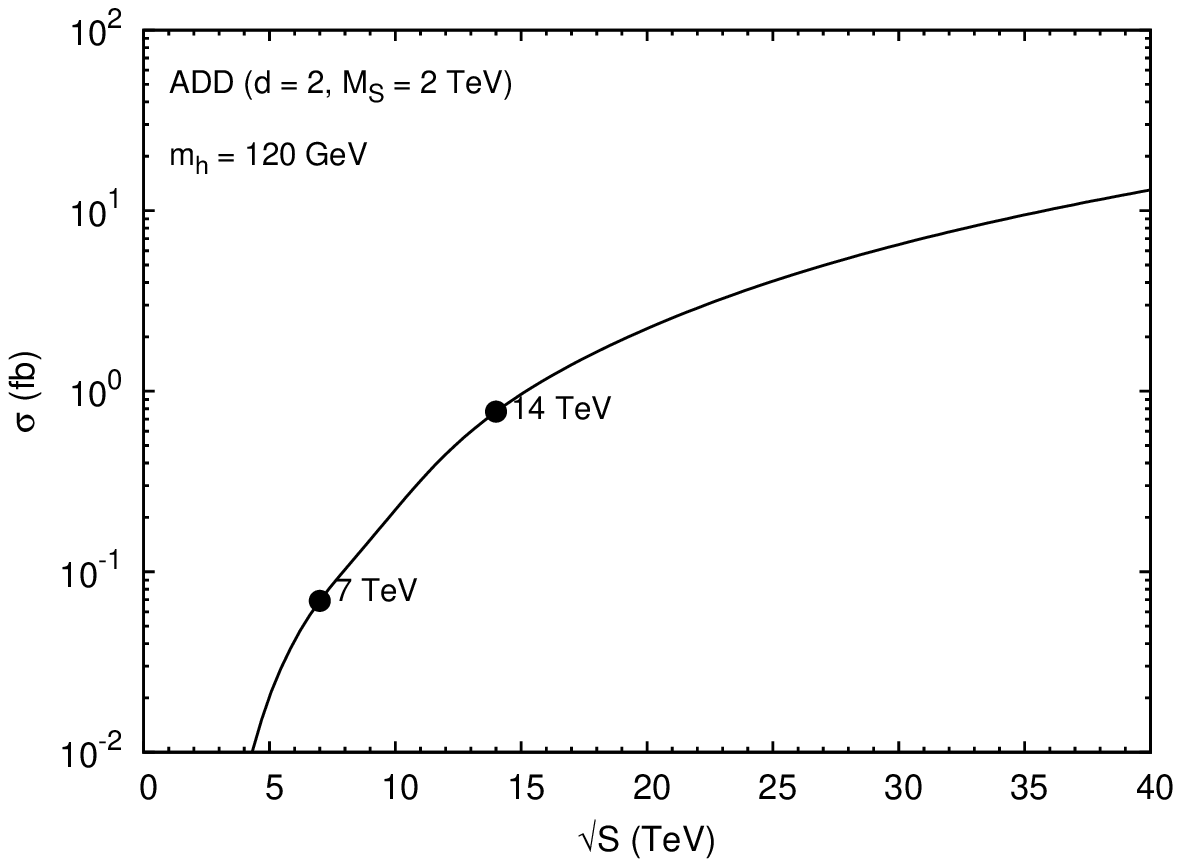}
\ec
\caption{Dependence of the cross section ($\s$) for the ADD model on the center of mass energy of the collider. For this plot 
 $M_{\rm S}$ = 2 TeV, $d = 2$ and $m_h$ = 120 GeV.}
\label{fig:E-X}
 \end{figure}

In Fig. \ref{fig:mh-X},  we show how the cross section changes with the Higgs boson mass, $m_h$. Here the cutoff scale, $M_{\rm S}$ 
has been set to 2 TeV and the number of compactified extra dimensions is 2. 
The cross section decreases with increasing Higgs boson mass primarily due to 
phase-space suppression. We see that over most of the plotted range, the cross section varies from about $0.9$ to $0.5$ fb. 
From this figure we also see that
the bottom quark loop contribution to the cross-section is negligible -- less than a percent (for the case of $gg\to h$ in the SM see 
\cite{Anastasiou:2009kn}). In Fig. \ref{fig:ms-X}, we show the 
 dependence of the cross section on the scale, $M_{\rm S}$ for different values of $d$. 
In Fig. \ref{fig:pt-X}, we have plotted the transverse momentum distribution of the Higgs boson for $d=2$ and  $M_{\rm S}$ = 2 TeV. 
We see that it has a peak at about 120 GeV. In Fig. \ref{fig:mkk-X}, we have plotted the dependence of  the cross section
on the $G_{\rm KK}$ mass for different values of $M_{\rm S}$ and $d$. Since the density of graviton states increases with the 
increase in the mass of the graviton, the cross section gets contribution from mostly large values of the  $G_{\rm KK}$ mass. For example 
for $d=2$ the peaks of the curves lie around 400 GeV (see Figs. \ref{fig:mkk-Xa} and \ref{fig:mkk-Xb}). However at the end 
phase space suppression takes over and the
cross section starts to decrease. From Eq. \ref{eq:dnsty_states} one would also expect 
this value to increase with the increase in $d$. This is seen in Figs. \ref{fig:mkk-Xc} and \ref{fig:mkk-Xd}, where for 
$d=4$ the peaks lie around 1 TeV. In Fig. \ref{fig:E-X} we show how the cross section increases with increasing
center of mass energy of the collider.

We see that for the parameter ranges considered the cross section is much smaller than what one would roughly estimate. To see that let us consider 
the SM process $gg\to h$ via a top quark loop (LO).
For 14 TeV LHC and $m_h\sim 120$ GeV the cross section of this process is about $20-25$ pb \cite{Djouadi:2005gi}. If one ignores the phase space suppression
due to the extra graviton, one would roughly expect the $gg \to h G_{\rm KK}$ process to be suppressed compared to the
$gg\to h$ process by an extra factor of $(E/M_{\rm S})^{(d+2)}$ where $E$ is the energy scale of the process. From Fig. \ref{fig:mkk-Xa}
we see that the maximum contribution comes from KK modes with mass $\sim$ 400 GeV. Hence if one takes $E\approx\sqrt{\hat s}\approx 500$ GeV,
then for $M_{\rm S} = 2$ TeV and $d=2$, one would expect the cross section to be about 80 fb compared to 0.8 fb that we get (see Fig. \ref{fig:mh-X}).
This happens because of the destructive interference between the contributions coming from the box-diagrams  and
the triangle-diagrams which happens because of the relative negative sign between these two sets of diagrams.
This cancellation reduces the amplitude drastically. For example, for $d=2$, $M_{\rm S}=2$ TeV and $m_h=120$ GeV, 
switching off the box diagrams leads to a cross section of 158805.5 fb \footnote{We quote this number just to demonstrate the large cancellation 
between the box and triangle contributions. However, one should keep in mind that the contribution coming only from the triangle diagrams is not gauge invariant.
} -- {\it i.e..} a increase of six orders of magnitude in the cross section. 
This roughly translates in to a cancellation of about two or three orders of magnitude at the amplitude level.
This cancellation and the cuts applied on physical quantities   
reduces the cross section to such small values.\footnote{Similar cancellation is also observed for the $gg\to Z G_{\rm KK}$ process\cite{ambresh}.} 
Still, one would expect a few hundred 
such events after the LHC achieves its design luminosity.

\begin{figure}[!ht]
\bc
\includegraphics [angle=0,width=.65\linewidth] {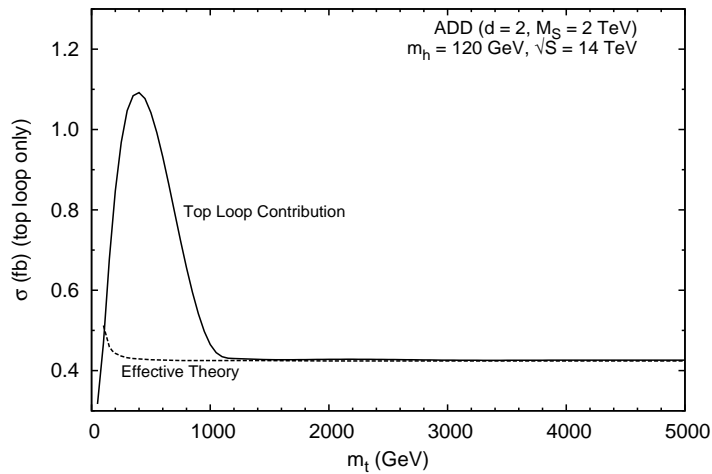}
\ec
\caption{Variation of the cross section ($\s$) with $m_t$ in the ADD model. 
For this plot the cutoff scale, $M_{\rm S} = 2$ TeV and the number of extra dimensions, $d = 2$. 
The effective theory contribution is also shown.}
\label{fig:mt-X}
\end{figure}
\begin{figure}[!h]\bc
\includegraphics [angle=0,width=.65\linewidth] {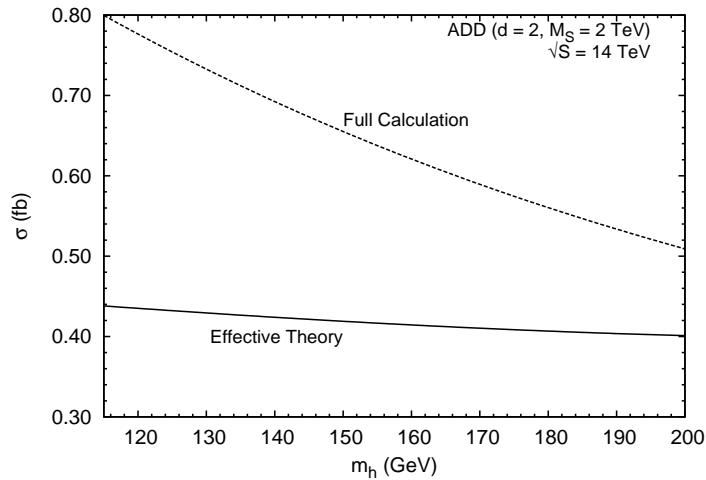}
\caption{Comparison between the effective theory and the full calculation. Here $m_t = 175$ GeV.}
\label{fig:mheft-X}\ec
 \end{figure}

Another interesting feature that we can see is that there is no decoupling of the top quark in the loop\cite{Appelquist:1974tg}. 
For this purpose, we have varied $m_t$ from 50 GeV to 5 TeV. From Fig. \ref{fig:mt-X} we clearly see that in the beginning the 
cross section increases due to the propagator enhancement. However, beyond $m_t \approx 400$ GeV, cross section  
decreases and approaches a constant value beyond $m_t \gtrsim 2$ TeV. This behavior
is similar to what has been seen in the case of $gg \to h$ production within the SM. In the large
$m_t$ limit, there also cross section is independent of $m_t$\cite{Georgi:1977gs}. In the SM,
because of the Higgs mechanism, we don't necessarily expect complete decoupling of the heavy
quarks, as it exists in the case of QED and QCD. In Fig. \ref{fig:mt-X} we also show the cross section computed using 
the effective theory approximation as described in Sec. \ref{subsec:heft}. In the numerical computation we keep upto $\mc O(\tau_h^3)$ 
terms which account for the $m_t$ dependence of the cross section for small $m_t$. The two calculations agree very well for $m_t \geq 1.2$ TeV. 
However for $m_h=120$ GeV and the physical top quark mass these two
differ. This can also be seen from Fig. \ref{fig:mheft-X}. This is unlike the SM $gg\to h$ case where this approximation  works 
quite well \cite{Pak:2009dg}. However in our case, there is 
one extra scale present -- namely, the mass of the graviton, $M_{\rm KK}$. We have seen that the  $M_{\rm KK}$ value is
significantly larger than the mass of the top quark most of the time. Because of this, $\sqrt{\hat s}$, which
is larger than $m_h + M_{\rm KK}$, can go
much beyond $2m_t$. Therefore,  one cannot expect the effective theory calculation to
agree with the full calculation for $m_h=120$ GeV and the physical top quark mass.

By using different CTEQ parton distributions CTEQ6M, CTEQ6D or CTEQ6L, 
we find
that the cross section can change by $5-15$ percent. We have also varied the factorization/renormalization
scale by a factor of two. We find that the cross section can vary by $20-25$ percent. This variation
can only be reduced by computing radiative correction to this process.
For $d = 2$ and $M_{\rm S} = 2$ TeV, the difference in the truncated and untruncated scheme cross sections
is about $15\%$ for $m_h = 120$ GeV. This difference increases to about $60\%$ for $d = 4$.
Furthermore, this difference keeps decreasing with the increase in the cut-off scale $M_{\rm S}$. These
results are consistent with the observation made in Refs.\cite{Karg:2009xk,Gao:2009pn}. Our results
will be modified if we include the QCD corrections to this process. Our process shares many features
with the process $gg \to h$. Therefore one may expect significant QCD corrections, i.e, a K-factor of
the order of 2.

\begin{figure}[t]
\bc
\includegraphics [angle=0,width=.65\linewidth] {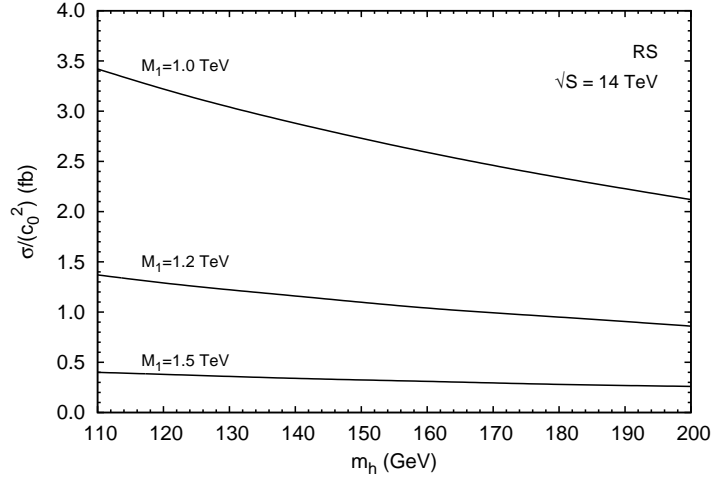}
\ec
\caption{Variation of the cross section ($\s$) scaled with the square of the dimensionless 
coupling $c_0=k/{\bar M_{\rm PL}}$ with the mass of the Higgs ($m_h$) for different values of $M_1$ -- 
mass of the first KK mode of the graviton -- in the RS model.}
\label{fig:rs-X}
\end{figure}
We have also computed the cross section for the RS model. In Fig. \ref{fig:rs-X}, we have plotted the scaled cross section
for three different values of the mass of the first KK mode of the graviton.
We see that, for $ 0.01 \lesssim c_0 \lesssim 0.1$, the cross section is more than an
order of magnitude smaller than that in the ADD model. For example for $c_0 = 0.075$, $M_1=1$ TeV and $m_h =120$ GeV the
cross section is only about 0.02 fb. The cross section 
is only significant for the smaller value of $M_1$ (correspondingly $m_0$) and larger value of $c_0$.

The question arises now -- can this process be observed at the LHC (assuming that both particles exist)?
To answer this one needs to understand the background. However, a complete analysis of the backgrounds is 
beyond the scope of this paper. 
We would discuss the possible signatures of this process semi-quantitatively. For $M_{\rm S} = 2$
TeV and $d = 2$, the cross section is of the order of a fb for the mass range $m_h = 110 - 200$ GeV.
In this mass range, the major decay modes of the Higgs boson are $h \to b {\bar b}, WW, ZZ$. We
will also take that graviton is not observed and it gives rise to a large missing $p_{\rm T}$ which can be used 
as a possible discriminator between the signal and the background. 
This can be seen from the the $p_{\rm T}$ distributions, shown in Fig. \ref{fig:pt-X}. 
For our process, the $p_{\rm T}$ of the graviton is same as that for the Higgs by the momentum conservation.

Let us now consider the signal and the backgrounds from the SM process for various major decay modes 
of the Higgs boson.
\begin{enumerate}
\item {\it $h \to b {\bar b}$}: This decay mode is dominant up-to about $m_h \approx 135$ GeV.
In the case of this decay mode, the signal will be ``two-$b$ jets + large missing $p_{\rm T}$''.
The main source of the background would be the production of ZZ pair and $Z + b {\bar b}$.
The typical cross section for $pp \to ZZX$ is about 10 pb. This is about four-orders of magnitude
higher than the signal cross section. The cross section for $Z + b {\bar b}$ production, with $Z$-boson
decaying into neutrinos, is also of the order of $10$ pb. One can suppress the backgrounds by considering the large
missing $p_{\rm T}$ and demanding the mass $M(b \bar{b})$ to be around the mass of the Higgs boson.
With the branching ratios of the $Z$ decays, we may be able to gain about
two-three orders of magnitude  in the signal-to-background ratio. Because of the small cross sections,
it may require many years of LHC run before this process could be seen through this decay 
mode of the Higgs boson.

\item {\it $h \to W^{+} W^{-}$}: This decay mode is important for $m_h \gtrsim 135$ GeV. 
The signature of this process can be ``two-leptons and large missing $p_{\rm T}$'' or ``one-lepton + 2 jets + large
missing $p_{\rm T}$''. The main sources of the backgrounds would be the production of WWZ and ZZ bosons.
The typical cross section for the background $pp \to WWZX$ is about $100$ fb. The ZZ background 
can be suppressed as the lepton pair from the Z-decay will have mass around $M_{\rm Z}$, while the lepton-pair from the
signal will have continuum distribution. The other background WWZ production can also be reduced
using larger missing $p_{\rm T}$ cut and the difference in $M(\ell \ell)$ distributions. Observation of this decay mode
may also take several years.

\item {\it $h \to Z Z$}: For $m_h \gtrsim 175$ GeV, this is the most important decay mode.
   For the SM Higgs boson production $gg \to h$, this decay mode
   gives rise to gold-plated signature of the ``two-lepton pairs''. In our case,
   this decay mode will give rise to two Z-bosons and large missing $p_{\rm T}$
          in the final state. The signal could be ``four-lepton + large missing $p_{\rm T}$'', or `` two leptons + two jets + large
          missing $p_{\rm T}$''. The main background is $pp \to ZZZX$  process. The typical cross section for this
          process is about 11 fb. To get large missing $p_{\rm T}$, one of the Z-boson will have to decay into neutrinos which
          has a branching ratio of about $20\%$. Using the large missing $p_{\rm T}$ and the mass of the two-lepton
          pairs (i.e. four leptons), this decay channel will give rise to the best observable signature
          of the process $gg \to h G_{\rm KK}$.
\end{enumerate}

\section{Conclusions}\label{sec:conclusions}

     In this paper, we have computed cross section and distributions for the process $pp \to hG_{\rm KK} \,X$
     for $m_h = 110 - 200$ GeV.
     This process occurs at the one-loop level through gluon-gluon fusion $gg \to hG_{\rm KK}$. The loop
     diagrams have been computed using the Oldenborgh-Vermaseren tensor-integral reduction procedure.
     We have presented the calculation in the ADD model, with a brief discussion in the context of the RS model.
     In the case of the ADD model, the cross section can be of the order of one fb. These values are
     smaller than expected due to destructive interference between the box-class and triangle-class
     of diagrams. This destructive interference reduces the amplitude by about two-orders of magnitude.
     The contribution to the cross section is mostly from large values of 
     $M_{\rm KK}$.
     We also note that the top quark does not decouple in the heavy top quark mass limit.
     We have also performed the computation using the effective theory approach. Only for very large top quark mass 
     this result matches very well with the exact calculation. 
     We find that the cross sections in the RS model are quite small. For $c_0 = 0.075$ and $M_1 = 1$ TeV,
      the cross section is about $0.02$ fb. We have also briefly considered possible 
      signatures of this process. It appears that for larger $m_h$, in the case of the ADD model,
      one may be able to observe this process at the LHC after a run of a few years.

\bigskip
\noi {\large\sf\bfseries Acknowledgments}\\ 
AS wants to thank M. Serone and I. Rothstein for fruitful discussions.

\setcounter{section}{0}
\renewcommand\thesection{Appendix \Alph{section}} 
\renewcommand{\theequation}{\Alph{section}.\arabic{equation}}
\newcommand{\appsection}[1]{\setcounter{equation}{0}\setcounter{figure}{0}\setcounter{table}{0}\section{#1}}

\appsection{Graviton Vertex Factors}\label{sec:notation}
The graviton vertex factors used in the calculation are
\ba
 V_{\rm qqG}^{\al\bt}(k_1,k_2) &=& \g^\al(k_1^{\bt} + k_2^{\bt}) + \g^\bt(k_1^{\al} + k_2^{\al}),\\
 V_{\rm hhG}^{\al\bt}(k_1,k_2) &=&  C^{\al\bt;\rho\s}\,k_{1\rho} k_{2\s}, \\
 V_{\rm ggG}^{\al\bt;\rho\s}(k_1,k_2) &=&
(k_1\cdot k_2)C^{\al\bt;\rho\s} + D^{\al\bt;\rho\s}(k_1,k_2) + \frac1\xi E^{\al\bt;\rho\s}(k_1,k_2), \\
 V_{\rm qqgG}^{\al\bt;\rho} &=&  C^{\al\bt;\rho\s}\,\g_\s.
\ea
For our computation, we have used the Feynman gauge, {\it i.e.}, $\xi=1$. 
The definitions of the functions $C$, $D$ and $E$ can be found in Ref. \cite{Han:1998sg}, which are
\ba
C^{\al\bt;\rho\sigma} &=& \eta^{\al\rho}\eta^{\bt\sigma}
+\eta^{\al\sigma}\eta^{\bt\rho}
-\eta^{\al\bt}\eta^{\rho\sigma},\\
D^{\al\bt;\rho\sigma} (k_1, k_2) &=&
\eta^{\al\bt} k_{1}^{\sigma}k_{2}^\rho
- \biggl[\eta^{\al\sigma} k_{1}^{\bt} k_{2}^{\rho}
  + \eta^{\al\rho} k_{1}^{\sigma} k_{2}^{\bt}
  - \eta^{\rho\sigma} k_{1}^{\al} k_{2}^{\bt}
  + (\al\leftrightarrow\bt)\biggr],\quad\\
E^{\al\bt;\rho\sigma} (k_1, k_2)&=& \eta^{\al\bt}(k_{1}^{\rho}k_{1}^{\sigma}
+k_{2}^{\rho}k_{2}^{\sigma}+k_{1}^{\rho}k_{2}^{\sigma})
\nonumber\\
&&\quad\quad\quad\quad
-\biggl[\eta^{\bt\sigma}k_{1}^{\al}k_{1}^{\rho}
+\eta^{\bt\rho}k_{2}^{\al}k_{2}^{\sigma}
+(\al\leftrightarrow\bt)\biggr].
\ea

\appsection{The $hqqG_{\rm KK}$ Vertex}\label{sec:newfeynrule}
The $hqqG^{\vec n}_{\rm KK}$ vertex, proportional to the Yukawa coupling ($y_q$) for a quark $q$, can be written as:
\bc
\includegraphics [angle=0,width=.30\linewidth] {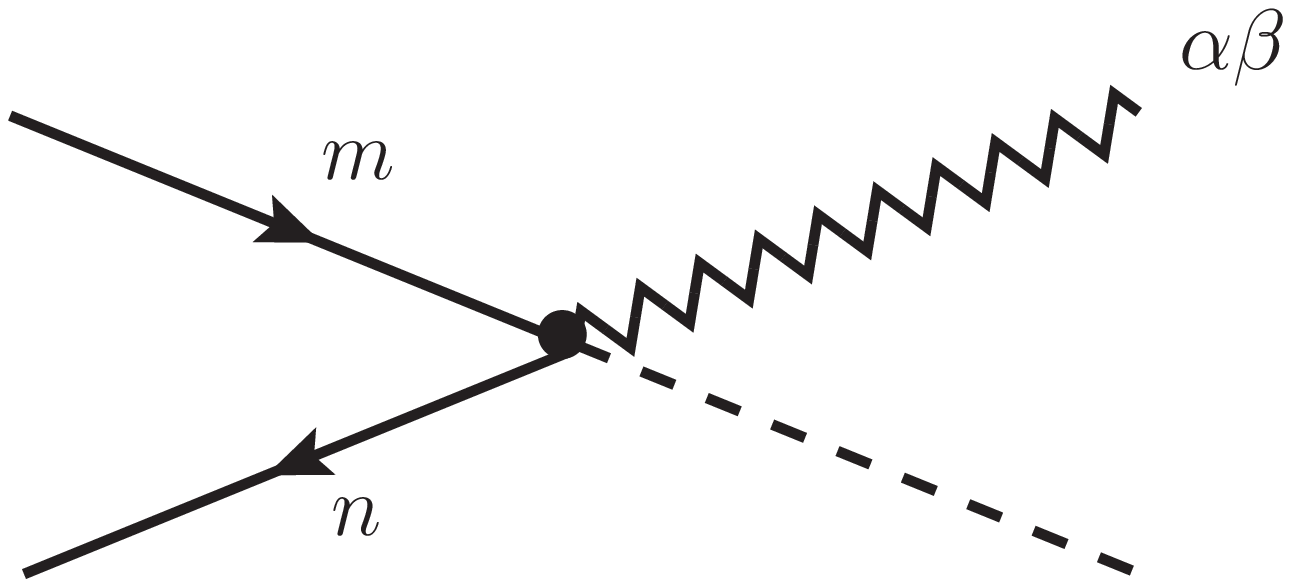}\label{fig:feynrules}
\ec\ba
\tilde h_{\al\bt}^{\vec n}\bar\psi\psi \Ph &:& \frac{i}2\, y_q\, \kp\, \et_{\al\bt}\,\dl_{mn}\\
\tilde \ph_{ij}^{\vec n}\bar\psi\psi \Ph &:& -2i\,\om\, y_q\, \kp\, \dl_{ij}\,\dl_{mn}
\ea
where we have followed the notation of Ref. \cite{Han:1998sg}. The Yukawa coupling is given as,
\ba
y_q &=& -\frac{g}{2m_W}\,m_q.
\ea

\appsection{The $hggG_{\rm KK}$ Vertex in the Effective theory}\label{sec:heftfeynrule}
The $hggG_{\rm KK}$ vertex is given by:
\bc
\includegraphics [angle=0,width=.30\linewidth] {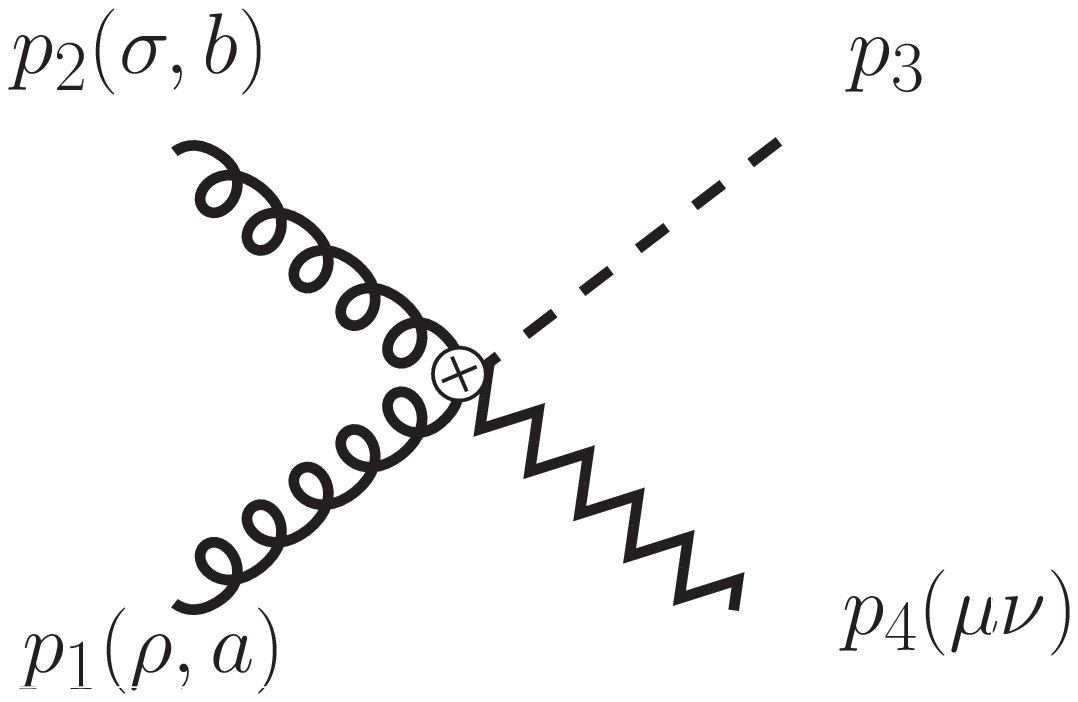}\label{fig:heftvert}
\ec\ba
\tilde h_{\m\n}^{\vec n} g g  \Ph &:& \frac{i\kp}2\, g_h \,\dl_{ab} \, \left[ \left(p_1\cdot p_2\right) C^{\m\n;\rho\sigma} 
+ D^{\m\n;\rho\sigma}(p_1,p_2)\right]\,.
\ea

\begin{spacing}{1}
\begin{small}

\end{small}
\end{spacing}

\begin{thebibliography}{unsrt}


\bibitem{Nakamura:2010zzi}
  K.~Nakamura {\it et al.}  [Particle Data Group],
  J.\ Phys.\ G {\bf 37}, 075021 (2010).


\bibitem{Lykken:2010mc}
  J.~D.~Lykken,
  arXiv:1005.1676 [hep-ph]~and the references therein.

\bibitem{jellis}
  J.~Ellis,
  arXiv:1102.5009 [hep-ph].

\bibitem{ADD}
  N.~Arkani-Hamed, S.~Dimopoulos and G.~R.~Dvali,
  Phys.\ Lett.\  B {\bf 429}, 263 (1998);
  [arXiv:hep-ph/9803315].

\bibitem{RS1}
  L.~Randall and R.~Sundrum,
  Phys.\ Rev.\ Lett.\  {\bf 83}, 3370 (1999);
  [arXiv:hep-ph/9905221].

\bibitem{RS2}
  L.~Randall and R.~Sundrum,
  Phys.\ Rev.\ Lett.\  {\bf 83}, 4690 (1999);
  [arXiv:hep-th/9906064].

\bibitem{Antoniadis}
  I.~Antoniadis,
  Phys.\ Lett.\  B {\bf 246}, 377 (1990).

\bibitem{Appelquist}
  T.~Appelquist, H.~C.~Cheng and B.~A.~Dobrescu,
  Phys.\ Rev.\  D {\bf 64}, 035002 (2001);
  [arXiv:hep-ph/0012100].

\bibitem{Davoudiasl:1999jd}
  H.~Davoudiasl, J.~L.~Hewett, T.~G.~Rizzo,
  Phys.\ Rev.\ Lett.\  {\bf 84}, 2080 (2000);
  [arXiv:hep-ph/9909255].

\bibitem{Han:1998sg}
  T.~Han, J.~D.~Lykken and R.~J.~Zhang,
  Phys.\ Rev.\  D {\bf 59}, 105006 (1999);
  [arXiv:hep-ph/9811350].

\bibitem{Giudice:1998ck}
  G.~F.~Giudice, R.~Rattazzi, J.~D.~Wells,
  Nucl.\ Phys.\  {\bf B544}, 3-38 (1999).
  [hep-ph/9811291].


\bibitem{Mirabelli:1998rt}
  E.~A.~Mirabelli, M.~Perelstein and M.~E.~Peskin,
  Phys.\ Rev.\ Lett.\  {\bf 82}, 2236 (1999)
  [arXiv:hep-ph/9811337].

\bibitem{Ghosh}
  D.~K.~Ghosh, S.~Raychaudhuri,
  Phys.\ Lett.\  {\bf B495}, 114-120 (2000);
  [arXiv:hep-ph/0007354].


\bibitem{Kaluza}
  T.~Kaluza,
  Sitzungsber.\ Preuss.\ Akad.\ Wiss.\ Berlin (Math.\ Phys.\ ) {\bf 1921}, 966 (1921).

\bibitem{Klein}
  O.~Klein,
  Z.\ Phys.\  {\bf 37}, 895 (1926)
  [Surveys High Energ.\ Phys.\  {\bf 5}, 241 (1986)].

\bibitem{Kapner}
  D.~J.~Kapner, T.~S.~Cook, E.~G.~Adelberger, J.~H.~Gundlach, B.~R.~Heckel, C.~D.~Hoyle and H.~E.~Swanson,
  Phys.\ Rev.\ Lett.\  {\bf 98}, 021101 (2007);
  [arXiv:hep-ph/0611184].

\bibitem{Chatrchyan}
  S.~Chatrchyan {\it et al.}  [CMS Collaboration],
  JHEP {\bf 1105}, 085 (2011);
  [arXiv:1103.4279 [hep-ex]].

\bibitem{Goldberger}
  W.~D.~Goldberger, M.~B.~Wise,
  Phys.\ Rev.\  {\bf D60}, 107505 (1999);
  [arXiv:hep-ph/9907218].

\bibitem{Mathews}
  P.~Mathews, V.~Ravindran, K.~Sridhar,
  JHEP {\bf 0510}, 031 (2005);
  [arXiv:hep-ph/0506158].

\bibitem{Aaltonen:2011xp}
  T.~Aaltonen {\it et al.} [CDF Collaboration],
  Phys.\ Rev.\ Lett.\  {\bf 107}, 051801 (2011);
  [arXiv:1103.4650 [hep-ex]].

\bibitem{Abazov:2010xh}
  V.~M.~Abazov {\it et al.}  [The D0 Collaboration],
  Phys.\ Rev.\ Lett.\  {\bf 104}, 241802 (2010);
  [arXiv:1004.1826 [hep-ex]].

\bibitem{Kumar:2010kv}
  M.~C.~Kumar, P.~Mathews, V.~Ravindran, S.~Seth,
  J.\ Phys.\ G {\bf G38}, 055001 (2011);
  [arXiv:1004.5519 [hep-ph]].

\bibitem{Kumar:2010ca}
  M.~C.~Kumar, P.~Mathews, V.~Ravindran, S.~Seth,
  Nucl.\ Phys.\  {\bf B847}, 54-92 (2011);
  [arXiv:1011.6199 [hep-ph]].

\bibitem{Karg:2009xk}
  S.~Karg, M.~Kramer, Q.~Li, D.~Zeppenfeld,
  Phys.\ Rev.\  {\bf D81}, 094036 (2010);
  [arXiv:0911.5095 [hep-ph]].

\bibitem{Gao:2009pn}
  X.~Gao, C.~S.~Li, J.~Gao, J.~Wang and R.~J.~Oakes,
  Phys.\ Rev.\  D {\bf 81}, 036008 (2010)
  [arXiv:0912.0199 [hep-ph]].

\bibitem{ambresh}
  A.~Shivaji, V.~Ravindran and P.~Agrawal,
  arXiv:1111.6479 [hep-ph].

\bibitem{Vermaseren:2000nd}
  J.~A.~M.~Vermaseren,
  arXiv:math-ph/0010025.

\bibitem{'tHooft:1978xw}
  G.~'t Hooft and M.~J.~G.~Veltman,
  Nucl.\ Phys.\  B {\bf 153}, 365 (1979).

\bibitem{vanOldenborgh:1989wn}
  G.~J.~van Oldenborgh and J.~A.~M.~Vermaseren,
  Z.\ Phys.\  C {\bf 46}, 425 (1990).

\bibitem{agrawal}
  P.~Agrawal and G.~Ladinsky,
  Phys.\ Rev.\  D {\bf 63}, 117504 (2001);
  [arXiv:hep-ph/0011346].

\bibitem{vanOldenborgh:1990yc}
  G.~J.~van Oldenborgh,
  Comput.\ Phys.\ Commun.\  {\bf 66}, 1 (1991).

\bibitem{Rizzo:1979mf} 
  T.~G.~Rizzo,
  Phys.\ Rev.\ D {\bf 22}, 178 (1980)
  [Addendum-ibid.\ D {\bf 22}, 1824 (1980)].

\bibitem{Pumplin:2002vw}
  J.~Pumplin, D.~R.~Stump, J.~Huston, H.~L.~Lai, P.~M.~Nadolsky and W.~K.~Tung,
  JHEP {\bf 0207}, 012 (2002);
  [arXiv:hep-ph/0201195].



\bibitem{Anastasiou:2009kn} 
  C.~Anastasiou, S.~Bucherer and Z.~Kunszt,
  JHEP {\bf 0910}, 068 (2009)
  [arXiv:0907.2362 [hep-ph]].


\bibitem{Alwall:2011uj} 
  J.~Alwall, M.~Herquet, F.~Maltoni, O.~Mattelaer and T.~Stelzer,
  JHEP {\bf 1106}, 128 (2011)
  [arXiv:1106.0522 [hep-ph]].

\bibitem{Djouadi:2005gi}
  A.~Djouadi,
  Phys.\ Rept.\  {\bf 457}, 1 (2008)
  [arXiv:hep-ph/0503172].

\bibitem{Appelquist:1974tg}
  T.~Appelquist and J.~Carazzone,
  Phys.\ Rev.\  D {\bf 11}, 2856 (1975).


\bibitem{Georgi:1977gs}
  H.~M.~Georgi, S.~L.~Glashow, M.~E.~Machacek and D.~V.~Nanopoulos,
  Phys.\ Rev.\ Lett.\  {\bf 40}, 692 (1978).


\bibitem{Pak:2009dg} 
  A.~Pak, M.~Rogal and M.~Steinhauser,
  JHEP {\bf 1002}, 025 (2010)
  [arXiv:0911.4662 [hep-ph]].




\end{thebibliography}
\end{document}